\documentclass[aps,prl,twocolumn,superscriptaddress]{revtex4-2}
\usepackage{color,xcolor}

\usepackage{footnote}
\usepackage[colorinlistoftodos,bordercolor=red!50,backgroundcolor=red!10,linecolor=red!50,textsize=tiny]{todonotes}
\usepackage{changes}

\usepackage[utf8]{inputenc}
\usepackage[T1]{fontenc}
\usepackage[english]{babel}  % English language
\usepackage{times}
\usepackage{bm}
\usepackage{ulem}

\usepackage[colorlinks=true,linkcolor=blue,citecolor=blue,urlcolor=blue,pdfauthor={ },pdftitle={ },pdfsubject={ },pdfkeywords={ }]{hyperref}
\usepackage{amssymb,amsmath,amsfonts,amsthm}
\usepackage{mathtools}
\usepackage{verbatim}
\usepackage{enumerate}
\usepackage{graphicx}
\usepackage{hyperref}
\usepackage{bbm}
\usepackage{booktabs}

\usepackage[caption=false]{subfig}

\usepackage{setspace}
\usepackage{array}

\usepackage{color}
\definecolor{dred}{rgb}{.8,0.2,.2}
\definecolor{ddred}{rgb}{.8,0.5,.5}
\definecolor{dblue}{rgb}{.2,0.2,.8}
\definecolor{dgreen}{rgb}{.2,0.5,.2}

\newcommand{\bra}[1]{\mbox{$\langle #1|$}}
\newcommand{\ket}[1]{\ensuremath{|#1\rangle}}

\newcommand{\be}{\begin{equation}}
\newcommand{\ee}{\end{equation}}
\newcommand{\bea}{\begin{eqnarray}}
\newcommand{\eea}{\end{eqnarray}}

\begin{document}

\title{Experimental sample-efficient quantum state tomography via parallel measurements}

\author{Chang-Kang Hu}
\thanks{These authors contributed equally to this work.}
\affiliation{International Quantum Academy, Shenzhen 518048, China}
\affiliation{Shenzhen Institute for Quantum Science and Engineering and Department of Physics, Southern University of Science and Technology, Shenzhen 518055, China}
\affiliation{Guangdong Provincial Key Laboratory of Quantum Science and Engineering,
Southern University of Science and Technology, Shenzhen 518055, China}
\affiliation{Shenzhen Key Laboratory of Quantum Science and Engineering, Southern University of Science and Technology, Shenzhen,518055, China}

\author{Chao Wei}
\thanks{These authors contributed equally to this work.}
\affiliation{Shenzhen Institute for Quantum Science and Engineering and Department of Physics, Southern University of Science and Technology, Shenzhen 518055, China}
\affiliation{Guangdong Provincial Key Laboratory of Quantum Science and Engineering,
Southern University of Science and Technology, Shenzhen 518055, China}
\affiliation{Shenzhen Key Laboratory of Quantum Science and Engineering, Southern University of Science and Technology, Shenzhen,518055, China}

\author{Chilong Liu}
\thanks{These authors contributed equally to this work.}
\affiliation{Shenzhen Institute for Quantum Science and Engineering and Department of Physics, Southern University of Science and Technology, Shenzhen 518055, China}
\affiliation{Guangdong Provincial Key Laboratory of Quantum Science and Engineering,
Southern University of Science and Technology, Shenzhen 518055, China}
\affiliation{Shenzhen Key Laboratory of Quantum Science and Engineering, Southern University of Science and Technology, Shenzhen,518055, China}

\author{Liangyu Che}
\affiliation{Shenzhen Institute for Quantum Science and Engineering and Department of Physics, Southern University of Science and Technology, Shenzhen 518055, China}
\affiliation{Guangdong Provincial Key Laboratory of Quantum Science and Engineering,
Southern University of Science and Technology, Shenzhen 518055, China}
\affiliation{Shenzhen Key Laboratory of Quantum Science and Engineering, Southern University of Science and Technology, Shenzhen,518055, China}

\author{Yuxuan Zhou}
\affiliation{Shenzhen Institute for Quantum Science and Engineering and Department of Physics, Southern University of Science and Technology, Shenzhen 518055, China}
\affiliation{Guangdong Provincial Key Laboratory of Quantum Science and Engineering,
Southern University of Science and Technology, Shenzhen 518055, China}
\affiliation{Shenzhen Key Laboratory of Quantum Science and Engineering, Southern University of Science and Technology, Shenzhen,518055, China}

\author{Guixu Xie}
\affiliation{Shenzhen Institute for Quantum Science and Engineering and Department of Physics, Southern University of Science and Technology, Shenzhen 518055, China}
\affiliation{Guangdong Provincial Key Laboratory of Quantum Science and Engineering,
Southern University of Science and Technology, Shenzhen 518055, China}
\affiliation{Shenzhen Key Laboratory of Quantum Science and Engineering, Southern University of Science and Technology, Shenzhen,518055, China}

\author{Haiyang Qin}
\affiliation{Shenzhen Institute for Quantum Science and Engineering and Department of Physics, Southern University of Science and Technology, Shenzhen 518055, China}
\affiliation{Guangdong Provincial Key Laboratory of Quantum Science and Engineering,
Southern University of Science and Technology, Shenzhen 518055, China}
\affiliation{Shenzhen Key Laboratory of Quantum Science and Engineering, Southern University of Science and Technology, Shenzhen,518055, China}
\author{Guantian Hu}
\affiliation{Shenzhen Institute for Quantum Science and Engineering and Department of Physics, Southern University of Science and Technology, Shenzhen 518055, China}
\affiliation{Guangdong Provincial Key Laboratory of Quantum Science and Engineering,
Southern University of Science and Technology, Shenzhen 518055, China}
\affiliation{Shenzhen Key Laboratory of Quantum Science and Engineering, Southern University of Science and Technology, Shenzhen,518055, China}

\author{Haolan Yuan}
\affiliation{Shenzhen Institute for Quantum Science and Engineering and Department of Physics, Southern University of Science and Technology, Shenzhen 518055, China}
\affiliation{Guangdong Provincial Key Laboratory of Quantum Science and Engineering,
Southern University of Science and Technology, Shenzhen 518055, China}
\affiliation{Shenzhen Key Laboratory of Quantum Science and Engineering, Southern University of Science and Technology, Shenzhen,518055, China}

\author{Ruiyang Zhou}
\affiliation{Shenzhen Institute for Quantum Science and Engineering and Department of Physics, Southern University of Science and Technology, Shenzhen 518055, China}
\affiliation{Guangdong Provincial Key Laboratory of Quantum Science and Engineering,
Southern University of Science and Technology, Shenzhen 518055, China}
\affiliation{Shenzhen Key Laboratory of Quantum Science and Engineering, Southern University of Science and Technology, Shenzhen,518055, China}

\author{Song Liu}
\email{lius3@sustech.edu.cn}
\affiliation{International Quantum Academy, Shenzhen 518048, China}
\affiliation{Shenzhen Institute for Quantum Science and Engineering and Department of Physics, Southern University of Science and Technology, Shenzhen 518055, China}
\affiliation{Guangdong Provincial Key Laboratory of Quantum Science and Engineering,
Southern University of Science and Technology, Shenzhen 518055, China}
\affiliation{Shenzhen Key Laboratory of Quantum Science and Engineering, Southern University of Science and Technology, Shenzhen,518055, China}
\affiliation{Shenzhen Branch, Hefei National Laboratory, Shenzhen 518048, China}

\author{Dian Tan}
\email{tand@sustech.edu.cn}
\affiliation{International Quantum Academy, Shenzhen 518048, China}
\affiliation{Shenzhen Institute for Quantum Science and Engineering and Department of Physics, Southern University of Science and Technology, Shenzhen 518055, China}
\affiliation{Guangdong Provincial Key Laboratory of Quantum Science and Engineering,
Southern University of Science and Technology, Shenzhen 518055, China}
\affiliation{Shenzhen Key Laboratory of Quantum Science and Engineering, Southern University of Science and Technology, Shenzhen,518055, China}

\author{Tao Xin}
\email{xint@sustech.edu.cn}
\affiliation{International Quantum Academy, Shenzhen 518048, China}
\affiliation{Shenzhen Institute for Quantum Science and Engineering and Department of Physics, Southern University of Science and Technology, Shenzhen 518055, China}
\affiliation{Guangdong Provincial Key Laboratory of Quantum Science and Engineering,
Southern University of Science and Technology, Shenzhen 518055, China}
\affiliation{Shenzhen Key Laboratory of Quantum Science and Engineering, Southern University of Science and Technology, Shenzhen,518055, China}

\author{Dapeng Yu}
\email{yudp@sustech.edu.cn}
\affiliation{International Quantum Academy, Shenzhen 518048, China}
\affiliation{Shenzhen Institute for Quantum Science and Engineering and Department of Physics, Southern University of Science and Technology, Shenzhen 518055, China}
\affiliation{Guangdong Provincial Key Laboratory of Quantum Science and Engineering,
Southern University of Science and Technology, Shenzhen 518055, China}
\affiliation{Shenzhen Key Laboratory of Quantum Science and Engineering, Southern University of Science and Technology, Shenzhen,518055, China}
\affiliation{Shenzhen Branch, Hefei National Laboratory, Shenzhen 518048, China}

\begin{abstract}
Quantum state tomography (QST) via local measurements on reduced density matrices (LQST) is a promising approach but becomes impractical for large systems. To tackle this challenge,  we developed an efficient quantum state tomography method inspired by quantum overlapping tomography [\href{10.1103/PhysRevLett.124.100401}{Phys. Rev. Lett. 124, 100401(2020)}], which utilizes parallel measurements (PQST). In contrast to LQST, PQST significantly reduces the number of measurements and offers more robustness against shot noise.  Experimentally, we demonstrate the feasibility of PQST in a tree-like superconducting qubit chip by designing high-efficiency circuits, preparing W states, ground states of Hamiltonians and random  states, and then reconstructing these density matrices using full quantum state tomography (FQST), LQST, and PQST. Our results show that PQST reduces  measurement cost, achieving fidelities of 98.68\% and 95.07\% after measuring 75 and 99 observables for 6-qubit and 9-qubit W states, respectively. Furthermore, the reconstruction of the largest density matrix of the 12-qubit W state is achieved with the similarity of 89.23\% after just measuring $243$ parallel observables, while $3^{12}=531441$  complete observables are needed for FQST. Consequently, PQST will be a useful tool for future tasks such as the reconstruction, characterization, benchmarking, and properties learning of states.
\end{abstract}

\maketitle
%%%%%%%%%%%%%%%%%%%%%%%%%%%%%%%%%%%%%%%%%%%%%%%%%%%%%
\textit{Introduction.—} 
Quantum state tomography is essential for characterizing quantum systems and enabling precise state reconstruction, which is critical for quantum information science \cite{paris2004quantum, Lvovsky2009}. The extensive interest in QST has grown due to  its applications in entanglement sources \cite{dunn1995experimental, sanaka2001new}, nonhermitian physics \cite{naghiloo2019quantum}, quantum teleportation \cite{bao2012quantum}, and nanoelectronics \cite{bisognin2019quantum}. Currently, a contradiction arises as quantum devices have been substantially scaled up to seek quantum advantages, yet extracting results has become increasingly challenging. This can potentially undermine promised speedup, because FQST containing exponentially increasing data collection and postprocessing, is unrealistic for large systems. A huge gap remains between the abilities to build quantum devices and to reconstruct their density matrices for full information of the state \cite{gebhart2023learning,eisert2020quantum,cruz2019efficient,kliesch2021theory}. For instance, performing FQST on a 10-qubit state can take around five days \cite{song201710}, and the time required grows dramatically as the number of qubits increases \cite{arute2019quantum, kim2023evidence, cao2023generation,patra2023efficient,ebadi2021quantum,bao2024schr,moses2023race}.

To remove this daunting bottleneck, a number of solutions were proposed \cite{gross2010quantum,riofrio2017experimental,flammia2011direct,cramer2010efficient,torlai2018neural,zhang2024almost,chai2023multiqubit,xin2017quantum,xin2019local}, such as QST via compressing sensing \cite{gross2010quantum,riofrio2017experimental}, neural network  \cite{torlai2018neural}, and reduced density matrices (RDMs) \cite{xin2017quantum,xin2019local}. Among these, LQST stands out because it determines the $N$-qubit states with $ O(3^kN^k)$ local observables involved in all $k$-qubit RDMs, making it experimentally feasible \cite{zhang2024almost, xin2017quantum,linden2002almost,parashar2009n,tyc2015quantum}. However, LQST still consumes lots of data collection time for large systems. In fact, because different RDMs overlap, measurements on one RDM also provide information about other overlapping RDMs.  By efficiently organizing overlapping information, the measurement cost can be reduced. Notably, J. Cotler and F. Wilczek proposed quantum overlapping tomography (QOT), which enables the determination of all $k$-qubit RDMs through logarithmic parallel measurements \cite{cotler2020quantum}, an approach that was independently proposed by X. Bonet-Monroig {\emph{et al.}} as well \cite{bonet2020nearly}. A recent work attempted to reconstruct global states from RDMs using parallel measurements, but it only reconstructs two-qubit RDMs, partially reflecting QOT by focusing on easily obtained nearest-neighbor two-qubit RDMs \cite{guo2023scalable}.  QOT can be applied to performing state tomography \cite{xin2017quantum}, measuring quantum correlations \cite{carleo2017solving}, classifying topological orders \cite{levin2006detecting}, and determining two-qubit RDMs in optical platforms \cite{yang2023experimental}. However, the feasibility of applying this technique in full-state tomography remains an unexplored problem.

In this Letter, we develop a superconducting qubit chip and design high-efficiency circuits for preparing W states \cite{dur2000three, acin2001classification}, ground states \cite{tilly2022variational}, and random states \cite{fisher2023random}. By using parallel measurements, we achieve higher fidelities in the sample-efficient reconstruction of their density matrices compared to LQST. Furthermore, we reconstruct the largest density matrix of 12-qubit W state after a  few minutes of data collection, whereas FQST would require over sixty days. Our work fully supports the practical application of QOT in state reconstruction.

 \begin{figure}
 \centering
  \includegraphics[width=0.43\textwidth]{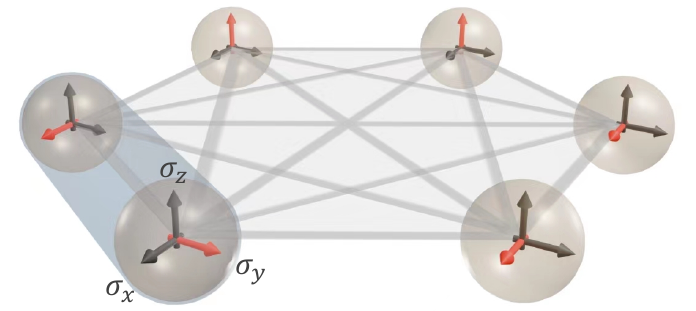}
   \caption{ The demonstration for 6-qubit QST. FQST requires measuring each qubit in the basis $\sigma_{x,y,z}$, resulting in $3^N=729$ global observables being measured. LQST requires measuring ${N\choose k}=15$ two-qubit RDMs (the gray lines), with $3^k=9$ local observables measured for each two-qubit RDM (indicated using a blue capsule), in total 135 local observables. Using PQST, all two-qubit RDMs are efficiently obtained through measuring 21 parallel observables, with one example highlighted by the red arrows. } 
  \label{fig11}
\end{figure}

%originating statistical distributions in the eigenvectors of Measurement data on a global observable $\mathcal{G}$ can infer expectation values of any local observables that overlap with $\mathcal{G}$. For instance, measuring $\sigma^{(1)}_x\otimes\sigma^{(2)}_y\otimes\sigma^{(3)}_z$ yields the expectation values of the one-qubit observables $\sigma^{(1)}_x, \sigma^{(2)}_y, \sigma^{(3)}_z$ and the two-qubit observables $\sigma^{(1)}_x\otimes\sigma^{(2)}_y, \sigma^{(2)}_y\otimes\sigma^{(3)}_z, \sigma^{(1)}_x\otimes\sigma^{(3)}_z$.

\begin{figure*}[tbh]
    \centering
  \includegraphics[width=0.96\textwidth]{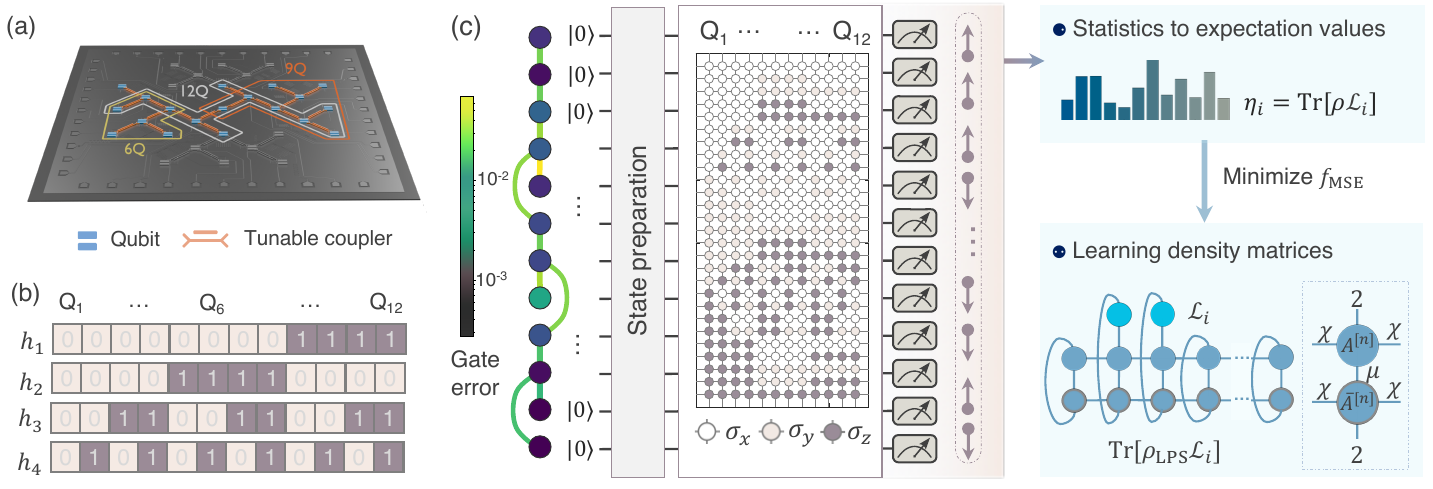}
    \caption{Preparation and the density matrix reconstruction of 12-qubit W state in a tree-like superconducting qubit chip. (a) The chip contains 31 transmon qubits, from which we choose 6-, 9-, 12-qubit structure for our experiments. (b) Design parallel observables by taking  $N=12$ with $k=2$ as an example.  There are four hash functions $h_{1,2,3,4}$ where the qubits with $h_i(j)=0$ (light color) or $h_i(j)=1$ (dark color) are represented. This configuration enables the determination of 27 parallel observables, as shown in the middle part of subfigure (c). (c) The entire scheme includes state preparation, parallel measurement, and density matrix learning using LPS \cite{sm}.  } 
    \label{fign2}
\end{figure*}

\textit{Parallel-measurement-based QST.—} Given an  $N$-qubit density matrix $\rho$, it can be expanded as,
\begin{equation}
\rho=\frac{1}{2^N}\sum_{i,j,...,l=0}^3\pi_{ij...l}\sigma^{(1)}_i\otimes \sigma^{(2)}_j ... \otimes \sigma^{(N)}_l,
\end{equation}
where $\sigma_0=\mathbbm{1}$ and $\sigma_{1,2,3}=\sigma_{x,y,z}$ represent the Pauli matrices. $\pi_{00...0}=1$ due to unit trace. To reconstruct  $\rho$, one usually create many samples of $\rho$ and measure expectation values $\pi_{ij...l}$ of $4^N-1$ Pauli observables $\sigma^{(1)}_i\otimes \sigma^{(2)}_j ... \otimes \sigma^{(N)}_{l}$.  Since the expectation values of local observables (with $\sigma_0$ in qubits) can be derived from the measurements on global observables (without $\sigma_0$ in each qubit) \cite{sm},  FQST requires measuring $3^N$ global Pauli observables, which is unfeasible for large systems. Quantum states are typically well-determined by their $k$-qubit RDMs \cite{zhang2024almost,parashar2009n,qi2019determining}, allowing global states to be reconstructed from local measurements on these RDMs \cite{linden2002parts, xin2017quantum}. This approach is known as LQST. There are $N \choose k$ such RDMs, each reconstructed by measuring $3^k$ local observables. Thus, LQST requires measuring $D_{\text{LQST}}=3^k{N\choose k}$ local observables to reconstruct $\rho$. However, since all $k$-qubit RDMs overlap with each other, it is unnecessary to measure each RDM individually. Instead, measuring each qubit in parallel provides information on multiple RDMs simultaneously, thereby reducing the measurement cost for reconstructing all $k$-qubit RDMs. The approach of determining RDMs through parallel measurements and subsequently reconstructing the full states is called PQST. See Fig. \ref{fig11} for a demonstration. Next, we introduce the PQST process, comprising the following steps. \\
\textbf{(1)} Design parallel observables. This step aims to find a smaller set of global Pauli observables whose measurement data is sufficient to reconstruct all $k$-qubit RDMs.  Here, we explore the application of QOT by taking $k=2$ as an example, which is briefly summarized from  \cite{cotler2020quantum}. \\
\textbf{(1.1)} System division and coloring. The system is divided into two colors according to $q=\lceil \text{log}_2N \rceil$ strategies, with $\lceil \cdot \rceil$ for rounding up to the nearest integer. The $i$-th strategy corresponds to an $N$-dimensional hash function $h_i$, where $h_i(j)$ is $i$-th digit in the binary expansion of $(j-1)$ in a $q$-bit string \cite{fredman1984size}. In each $h_i$, the $j$-th qubit is assigned a light color if $h_i(j)=0$ and a dark color otherwise, as shown in Fig. \ref{fign2}(b). This ensures that at least one hash function assigns different colors to any two qubits.\\
\textbf{(1.2)} Arrange parallel observables. The qubits of the same color are measured in the same Pauli basis for each $h_i$, while different Pauli basis are arranged for qubits with different colors, resulting in six parallel observables for each $h_i$. Additionally, all the qubits are measured in the same Pauli basis, requiring three parallel observables. Therefore, a total of $D_{\text{PQST}}=3+6\lceil \text{log}_2N \rceil$ parallel observables are needed. For $k>2$,  
the hash functions can be found by transforming the problem into a clique cover and solving it via binary linear programming optimization \cite{sm}. This allows all local expectation values involved in all $k$-qubit RDMs to be efficiently determined by parallel measurements. \\
\textbf{(2) }Perform measurements. The projection measurements are implemented on each parallel observable. The measurement outcomes can be efficiently post-processed to obtain the expectation values of local observables (the $i$-th one is labeled by $\mathcal{L}_i$) involved in all $k$-qubit RDMs. These values are expressed as 
\begin{equation}
\eta_i\equiv\text{tr}(\mathcal{L}_i\rho)=\frac{1}{M_j}\sum_{s=1}^{M_j}\text{tr}(\ket{\gamma^j_s}\bra{\gamma^j_s}\cdot \mathcal{L}_i),
\end{equation}
Here, $M_j$ represents the number of measurement samples on the $j$-th parallel observable. $\ket{\gamma^j_s}$ denotes the outcome state of the $s$-th projection measurement on it. For example, $\ket{\gamma^j_s}=\ket{010}$ when measuring $\sigma^{(1)}_z\otimes\sigma^{(2)}_z\otimes\sigma^{(3)}_z$. \\
\textbf{(3)} Learning density matrices. $\rho$ is estimated by finding the state whose corresponding measurement results most closely match those of $\rho$. As illustrated in Fig. \ref{fign2}(c), we represent the density matrix using the locally purified state (LPS), denoted as  $\rho_{\text{LPS}}$, an extension of matrix product states that is suitable for mixed states and offers favorable complexity scaling with system size \cite{sharir2022neural,li2023efficient, deng2017quantum,schollwock2011density}. The Mean Square Error (MSE) \cite{prasad1990estimation} can serve as a loss function, defined as $f_{\text{MSE}}=\sum_{i=1}^S|\text{tr}(\rho_{\text{LPS}}\mathcal{L}_i)-\eta_i|^2/S$. $S$ is the number of observables involved in all $k$-qubit RDMs. As mentioned before, PQST uses parallel measurements to acquire these RDMs, resulting in more shots on local observable $\mathcal{L}_i$ and more precise estimation of $\eta_i$ than LQST under the same sample size. Moreover, PQST data provides additional information about $\rho$ beyond $k$-qubit RDMs. Measuring parallel observables also gives information on other observables not included in $k$-qubit RDMs. By directly incorporating PQST shot data into the negative logarithm of the Maximum Likelihood Estimation (MLE) loss function, $f_{\text{MLE}} = -\sum_{j=1}^{D_{\text{PQST}}}\sum_{s=1}^{M_j} \log_2 \text{tr}(\ket{\gamma^j_s}\bra{\gamma^j_s} \rho_{\text{LPS}})$, the QST performance is further enhanced. This is because $f_{\text{MLE}}$ inputs more information about $\rho$, whereas $f_{\text{MSE}}$ relies solely on the expectation values $\eta_i$ involved in $k$-qubit RDMs. The following experiments will validate these two insights.

\textit{Experiments.—} We employ a tree-like superconducting qubit chip to experimentally demonstrate the substantial advantage of the PQST applied to systems with up to 12 qubits.
Figure \ref{fign2}(a) depicts the chip design utilized in the experiment, featuring a flip-chip package with a top layer consisting of fixed-frequency qubits and adjustable-frequency couplers \cite{yan2018tunable}, where the couplers facilitate interactions between qubits. We cool the chip to a base temperature of around 10 mK in a dilution refrigerator to minimize thermal noise, enabling coherent qubit control and readout \cite{sm}. Setting suitable idling coupler frequencies helps mitigate unwanted coupling between qubits, ensuring high-fidelity single-qubit gates. Then, we use a 160 ns pulse to implement the adiabatic Controlled-Z  gates \cite{xu2020high}. Using cross-entropy benchmarking \cite{boixo2018characterizing}, we found the median single-qubit gate fidelity is 99.92\%, and the median two-qubit gate fidelity is 97.80\%. More details on the chip and the experimental setup can be found in \cite{sm}.

Figure \ref{fign2}(c) illustrates the experimental process. We prepare the W states by designing high-efficiency preparation circuits that maximize the parallelism of the two-qubit Controlled-Z gates and optimize the circuit performance based on the qubit connectivity and gate fidelity \cite{sm}. The W state (named after W. D{\"u}r \cite{dur2000three}), which represents multipartite entanglement as an equal-weight superposition of all terms with one qubit in $\ket{1}$ and all others  in $\ket{0}$, can be expressed as $\ket{W_N}=\frac{1}{\sqrt{N}}(\ket{00...01}+\ket{00...10}+...+\ket{10...00})$ \cite{eltschka2012entanglement}. Circuits for 6-, 9-, and 12-qubit W states are detailed in \cite{sm}. The circuit for preparing the 12-qubit W state is implemented as a 2.3 $\mu s$ long circuit, consisting of 112 single-qubit gates and 22 two-qubit Controlled-Z  gates. Then, we reconstruct these density matrices via FQST, LQST, and PQST, respectively. For FQST, a complete set of $3^N$ observables with $M=10^4$  samples for each observable are measured. This process takes approximately one hour for 6-qubit FQST and thirty hours for 9-qubit FQST. FQST serves as a reliable benchmark for comparing PQST reconstruction results. We skip 12-qubit FQST due to its excessive time cost.  For LQST, we measure  $D_{\text{LQST}}^{(12, 2)}=594$ local two-qubit observables to reconstruct all two-qubit RDMs and $D_{\text{LQST}}^{(12, 3)}=5940$ local three-qubit observables to reconstruct all three-qubit RDMs for 12-qubit. In contrast, using PQST, we only measure  $D_{\text{PQST}}^{(12, 2)}=27$ and $D_{\text{PQST}}^{(12, 3)}=243$ parallel observables. For each sample, we randomly select a measurement observable from these sets and perform a projection measurement in its eigenbasis, recording the "down" or "up" results for all qubits.  After that, we utilize LPS to efficiently learn these matrices, avoiding the large storage memory of mixed density matrices \cite{sm, lps}. 
 
 %The fundamental element of LPS is the four-bond tensor $A^{[n]}_{2\times\chi^2\times\mu}$, where the factors 2, $\chi$, and $\mu$ denote physical, ancillary and purification indexes, respectively. The number of LPS parameters scales favorably with the system size as $O(2N\chi^2\mu)$, avoiding the exponentially increase in state space and drastically reducing storage and learning costs. Experimentally, we choose $\chi=18$ and do not constrain pure states by setting each $\mu=2$.  Then tensors $A^{[n]}$ are updated via the gradient descent, using automatic differentiation to minimize the loss functions.

 \begin{figure*}[tbh]
    \centering
    \includegraphics[width=1\textwidth]{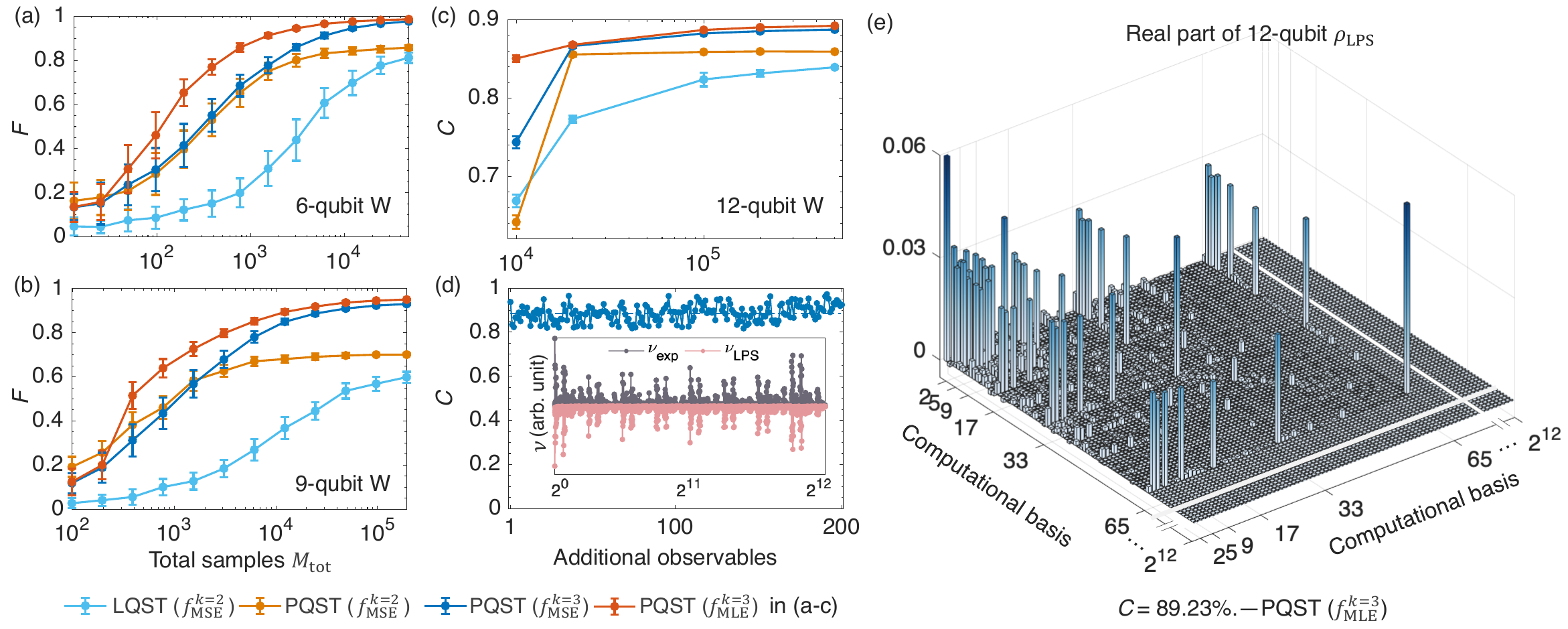}
    \caption{ (a-b) The fidelities of the reconstructed density matrices with those obtained from FQST under different $M_{\text{tot}}$ for 6-qubit and 9-qubit W states. The error bars are the estimated uncertainties from the repeated sampling, which primarily arises from fluctuations in the projection statistics due to the finite sample size \cite{bar}. This demonstrates that PQST is more robust against shot noise compared to LQST. (c) The similarity between the projection distribution vectors $\nu_{\text{LPS}}$ and $\nu_{\text{exp}}$ as a function of $M_{\text{tot}}$ for 12-qubit W state. (d)  The similarities of 200 additional observables corresponding to the last point on the red curve in subfigure (c). The size of the error bars is much smaller than the size of the data points. The insert shows an example comparing $\nu_{\text{LPS}}$ and $\nu_{\text{exp}}$ on one of these observables with error bars omitted due to the high dimensionality of up to $2^{12}$. Additional examples can be found in \cite{sm}. (e)  The reconstructed density matrix of the 12-qubit W state. The $2^{12}\times 2^{12}$ matrix is too large to be rendered entirely, so only the left and right part of the density matrix are shown.} 
    \label{fig22}
\end{figure*}

%The insert: the projection distribution $\nu_{\text{LPS}}$ of learned 12-qubit density matrix on one of these observables and its comparison with the experimental $\nu_{\text{exp}}$. 

\textit{Results.—} We have four main experimental results. First, we confirm the feasibility of QOT in state reconstruction and demonstrate the superiority of PQST over LQST in the sample size. Figures \ref{fig22}(a-b) present the fidelity $F=\text{tr}(\rho_{\text{LPS}}\rho_{\text{FQST}})/\sqrt{\text{tr}(\rho_{\text{LPS}}^2)\text{tr}(\rho_{\text{FQST}}^2)}$ between $\rho_{\text{LPS}}$ reconstructed via PQST or LQST and $\rho_{\text{FQST}}$ obtained via FQST \cite{Liang_2019}, plotted as a function of total samples $M_{\text{tot}}$ for 6-qubit and 9-qubit cases.  The fidelity reaches 98.68\% and 95.07\% for $6$- and $9$-qubit W states via PQST after just $5\times 10^4$ and $2\times 10^5$ samples, respectively, demonstrating that PQST is more sample-efficient. The fidelities of PQST and FQST with the theoretically-prepared states are provided in \cite{sm}, showing that the fidelity of PQST matches well with that of FQST and exhibits small fidelity uncertainties, despite PQST requiring far fewer measurement samples.

%The fidelities of FQST, LQST, and PQST with the theoretically-prepared states are provided in \cite{sm}, showing that the fidelity of PQST is closer to that of FQST, with smaller fidelity uncertainties compared to LQST under the same number of measurement samples
Second, we prepare and reconstruct the density matrix of the current largest W state with 12 qubits. Considering that 12-qubit FQST was not performed, we randomly measured 200 additional observables not included in PQST to verify the reconstruction accuracy. The projection distributions, representing the probability distribution of outcomes in the computational basis for each observable, were reshaped as vectors $\bf{\nu}$. Cosine similarity, measuring vector similarity \cite{han2012getting}, was used to evaluate the accuracy of the 12-qubit state reconstruction by comparing the projection distribution vectors $\bf{\nu}_{\text{exp}}$ from the experiment with those $\bf{\nu}_{\text{LPS}}$ from the reconstructed $\rho_{\text{LPS}}$. The similarity is defined as  $C=|\bf{\nu}_{\text{exp}}\cdot \bf{\nu}_{\text{LPS}}|/|\bf{\nu}_{\text{exp}}|/|\bf{\nu}_{\text{LPS}}|$ between $\bf{\nu}_{\text{exp}}$ and $\bf{\nu}_{\text{LPS}}$, ranging from 0 to 1, with values closer to 1 indicating higher similarity.  As shown in Fig. \ref{fig22}(c-d), PQST achieves the average similarity $C=$ 89.23\% after measuring $D_{\text{PQST}}^{(12,3)}=243$ parallel observables that cost $5\times 10^5$ samples (around four minutes of data collection), while FQST requires measuring $3^{12}=531441$ observables.

\begin{figure}[tbh]
  \centering
    \includegraphics[width=0.45\textwidth]{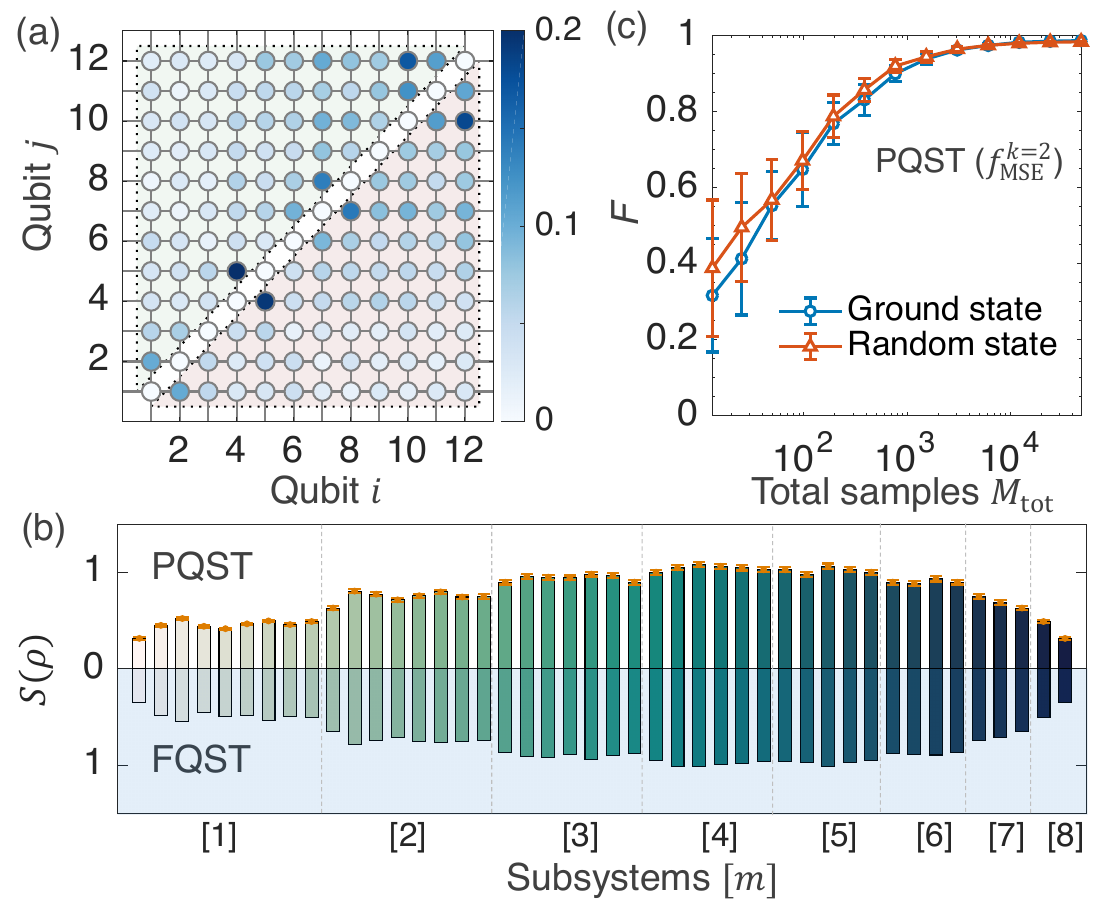}
   \caption{ (a) Experimental results of two-qubit correlations for a 12-qubit system using local (top left) and parallel (bottom right) measurements. The respective $M_{\text{tot}}$ are $594M$ and $27M$ with $M=500$, achieving over 95\% sample savings. The error bars are presented in \cite{sm}, showing that the uncertainty of parallel measurements is smaller. (b) Logarithmic negativity derived via FQST and PQST for the 9-qubit W state. $[m]$ denotes all nearest-neighbor subsystems with $m$ qubits \cite{bar}. 
  (c) The reconstruction results of 6-qubit ground states and random states using PQST with $f^{k=2}_\text{MSE}$.} 
  \label{fig44}
\end{figure}

Third, we demonstrate the strong power of parallel measurements in measuring multi-qubit correlators. There are $594$ two-qubit and $5940$ three-qubit correlators for 12-qubit, which can be obtained by only measuring  $27$ and $243$ parallel observables. In Fig. \ref{fig44}(a), as an example, we make the comparison of the correlation $\langle \sigma^{(i)}_y\sigma^{(j)}_y\rangle-\langle \sigma^{(i)}_y\rangle\langle\sigma^{(j)}_y\rangle$ obtained by local and parallel measurement methods. The good agreement suggests that parallel measurements provide an efficient means of measuring multi-qubit correlators. 

Finally, PQST measures far fewer observables than FQST to reconstruct the density matrix, but it enables predictions even on properties of unmeasured subsystems with comparable accuracy. Figure \ref{fig44}(b) presents the logarithmic negativity $S(\rho)=\text{log}_2||\rho^{\Gamma_A}||_1$ obtained via FQST and  PQST, where $\rho^{\Gamma_A}$ is the partial transpose of $\rho$ with respect to subsystem $A$ with $m$ qubits and $||\cdot||_1$ denotes the trace norm. Although PQST only measures $D_{\text{PQST}}^{(9,3)}=99$ parallel observables to reconstruct density matrices, it still has a good prediction even on the unmeasured subsystems with $m>3$, which agrees well with the result of FQST.

Moreover, to demonstrate the adaptability of PQST to various states, we experimentally prepare and reconstruct different types of states. 
For example, we prepare the ground state of a fully-connected Hamiltonian (with coupling between any two qubits) using a variational quantum eigensolver \cite{tilly2022variational}. Additionally, we prepare the final state of random circuits starting from $\ket{0}^{\otimes N}$ \cite{fisher2023random}.
We then reconstruct their density matrices with FQST, LQST, and PQST, considering two-qubit RDMs in LQST and PQST. In Fig.  \ref{fig44}(c), we plot $F(\rho_{\text{LPS}}, \rho_{\text{FQST}})$ as a function of  $M_{\text{tot}}$, demonstrating that PQST method accurately reconstructs these states (see LQST results in \cite{sm}). We also numerically demonstrate the reconstruction capability of PQST for dynamical states governed by fully-connected Hamiltonians. The results indicate that PQST accurately reconstructs these dynamical states at various evolution times, using $k=3$ for $N=6$ and $k=4$ for $N=9$ \cite{sm}.

\textit{Conclusions.—} In this work, we demonstrate a sample-efficient PQST that significantly reduces the measurement samples and offers greater robustness against shot noises compared to LQST in a superconducting qubit chip. Using parallel measurements and tensor network learning, we further achieve the largest density matrix reconstruction of the 12-qubit W state to date. Our work is the first to fully demonstrates the promising applications of  PQST for state reconstruction and state property learning, including two-qubit correlations and subsystem negativity \cite{lewis2024improved}, which holds significant value for experimentally characterizing many-body quantum states \cite{eisert2020quantum}, offering extensive applications in quantum chemistry and many-body physics simulations \cite{google2020hartree, arguello2019analogue,georgescu2014quantum}.

%We demonstrate the capability of PQST to generate 12-qubit entanglement experimentally by designing a high-efficiency circuit to prepare the 12-qubit W state.

\begin{acknowledgments}
\textit{Acknowledgments.—}
This work is supported by the Key-Area Research and Development Program of Guangdong Province (Grants No. 2018B030326001), the Guangdong Basic and Applied Basic Research Foundation (2022B1515020074), the National Natural Science Foundation of China (12275117, 12205137, 12004167, 11934010), the National Key Research and Development Program of China (2019YFA0308100), the China Postdoctoral Science Foundation (Grant No. 2020M671861, 2021T140648), the Guangdong Provincial Key Laboratory (Grant No.2019B121203002), Shenzhen Science and Technology Program (RCYX20200714114522109 and KQTD20200820113010023), Technology and Innovation Commission of Shenzhen Municipality (JCYJ20170412152620376, KYTDPT20181011104202253), and Center for Computational Science and Engineering at Southern University of Science and Technology. 
\end{acknowledgments}

%\bibliography{QOT.bib}

\begin{thebibliography}{62}
\bibitem [{lps()}]{lps}%
  \BibitemOpen
  \href@noop {} {\bibinfo  {journal} {The fundamental element of
  LPS is the four-bond tensor $A^{[n]}_{2\times\chi^2\times\mu}$, where the
  factors 2, $\chi$, and $\mu$ denote physical, ancillary and purification
  indexes, respectively. The number of LPS parameters scales favorably with the
  system size as $O(2N\chi^2\mu)$, avoiding the exponentially increase in state
  space and drastically reducing storage and learning costs. Experimentally, we
  choose $\chi=18$ and do not constrain pure states by setting each $\mu=2$.
  Then tensors $A^{[n]}$ are updated via the gradient descent, using automatic
  differentiation to minimize the loss functions}\ }\BibitemShut {NoStop}%
\bibitem [{bar()}]{bar}%
  \BibitemOpen
\bibfield  {journal} {  }\href@noop {} {\bibinfo  {journal} {To
  compute the uncertainties, we repeat the sampling experiments multiple times
  for each sample size. The fidelity uncertainty is then calculated using the
  standard deviation, defined as: $\Delta F = \sqrt{\sum_{i=1}^{R}(F_i -
  \bar{F})^2/{R}}$, where $F_i$ is the fidelity value of the $i$-th repeat
  experiment, and $\bar{F}$ is the mean fidelity of $R$ repeated experiments.
  Note that FQST experiments are conducted with a sufficient number of
  measurement samples, making the uncertainty from finite samples negligible.
  So we do not present error bars in the FQST results}\ }\BibitemShut
  {NoStop}%
\end{thebibliography}

%

\end{document}

\title{Supplemental Information for:\\ Experimental sample-efficient quantum state tomography via parallel measurements}

\author{Chang-Kang Hu}
\thanks{These authors contributed equally to this work.}
\affiliation{International Quantum Academy, Shenzhen 518048, China}
\affiliation{Shenzhen Institute for Quantum Science and Engineering and Department of Physics, Southern University of Science and Technology, Shenzhen 518055, China}
\affiliation{Guangdong Provincial Key Laboratory of Quantum Science and Engineering,
Southern University of Science and Technology, Shenzhen 518055, China}
\affiliation{Shenzhen Key Laboratory of Quantum Science and Engineering, Southern University of Science and Technology, Shenzhen,518055, China}

\author{Chao Wei}
\thanks{These authors contributed equally to this work.}
\affiliation{Shenzhen Institute for Quantum Science and Engineering and Department of Physics, Southern University of Science and Technology, Shenzhen 518055, China}
\affiliation{Guangdong Provincial Key Laboratory of Quantum Science and Engineering,
Southern University of Science and Technology, Shenzhen 518055, China}
\affiliation{Shenzhen Key Laboratory of Quantum Science and Engineering, Southern University of Science and Technology, Shenzhen,518055, China}

\author{Chi-Long Liu}
\thanks{These authors contributed equally to this work.}
\affiliation{Shenzhen Institute for Quantum Science and Engineering and Department of Physics, Southern University of Science and Technology, Shenzhen 518055, China}
\affiliation{Guangdong Provincial Key Laboratory of Quantum Science and Engineering,
Southern University of Science and Technology, Shenzhen 518055, China}
\affiliation{Shenzhen Key Laboratory of Quantum Science and Engineering, Southern University of Science and Technology, Shenzhen,518055, China}

\author{Liangyu Che}
\affiliation{Shenzhen Institute for Quantum Science and Engineering and Department of Physics, Southern University of Science and Technology, Shenzhen 518055, China}
\affiliation{Guangdong Provincial Key Laboratory of Quantum Science and Engineering,
Southern University of Science and Technology, Shenzhen 518055, China}
\affiliation{Shenzhen Key Laboratory of Quantum Science and Engineering, Southern University of Science and Technology, Shenzhen,518055, China}

\author{Yu-Xuan Zhou}
\affiliation{Shenzhen Institute for Quantum Science and Engineering and Department of Physics, Southern University of Science and Technology, Shenzhen 518055, China}
\affiliation{Guangdong Provincial Key Laboratory of Quantum Science and Engineering,
Southern University of Science and Technology, Shenzhen 518055, China}
\affiliation{Shenzhen Key Laboratory of Quantum Science and Engineering, Southern University of Science and Technology, Shenzhen,518055, China}

\author{Gui-Xu Xie}
\affiliation{Shenzhen Institute for Quantum Science and Engineering and Department of Physics, Southern University of Science and Technology, Shenzhen 518055, China}
\affiliation{Guangdong Provincial Key Laboratory of Quantum Science and Engineering,
Southern University of Science and Technology, Shenzhen 518055, China}
\affiliation{Shenzhen Key Laboratory of Quantum Science and Engineering, Southern University of Science and Technology, Shenzhen,518055, China}

\author{Haiyang Qin}
\affiliation{Shenzhen Institute for Quantum Science and Engineering and Department of Physics, Southern University of Science and Technology, Shenzhen 518055, China}
\affiliation{Guangdong Provincial Key Laboratory of Quantum Science and Engineering,
Southern University of Science and Technology, Shenzhen 518055, China}
\affiliation{Shenzhen Key Laboratory of Quantum Science and Engineering, Southern University of Science and Technology, Shenzhen,518055, China}
\author{Guan-Tian Hu}
\affiliation{Shenzhen Institute for Quantum Science and Engineering and Department of Physics, Southern University of Science and Technology, Shenzhen 518055, China}
\affiliation{Guangdong Provincial Key Laboratory of Quantum Science and Engineering,
Southern University of Science and Technology, Shenzhen 518055, China}
\affiliation{Shenzhen Key Laboratory of Quantum Science and Engineering, Southern University of Science and Technology, Shenzhen,518055, China}

\author{Hao-Lan Yuan}
\affiliation{Shenzhen Institute for Quantum Science and Engineering and Department of Physics, Southern University of Science and Technology, Shenzhen 518055, China}
\affiliation{Guangdong Provincial Key Laboratory of Quantum Science and Engineering,
Southern University of Science and Technology, Shenzhen 518055, China}
\affiliation{Shenzhen Key Laboratory of Quantum Science and Engineering, Southern University of Science and Technology, Shenzhen,518055, China}

\author{Rui-Yang Zhou}
\affiliation{Shenzhen Institute for Quantum Science and Engineering and Department of Physics, Southern University of Science and Technology, Shenzhen 518055, China}
\affiliation{Guangdong Provincial Key Laboratory of Quantum Science and Engineering,
Southern University of Science and Technology, Shenzhen 518055, China}
\affiliation{Shenzhen Key Laboratory of Quantum Science and Engineering, Southern University of Science and Technology, Shenzhen,518055, China}

\author{Song Liu}
\email{lius3@sustech.edu.cn}
\affiliation{International Quantum Academy, Shenzhen 518048, China}
\affiliation{Shenzhen Institute for Quantum Science and Engineering and Department of Physics, Southern University of Science and Technology, Shenzhen 518055, China}
\affiliation{Guangdong Provincial Key Laboratory of Quantum Science and Engineering,
Southern University of Science and Technology, Shenzhen 518055, China}
\affiliation{Shenzhen Key Laboratory of Quantum Science and Engineering, Southern University of Science and Technology, Shenzhen,518055, China}

\author{Dian Tan}
\email{tand@sustech.edu.cn}
\affiliation{International Quantum Academy, Shenzhen 518048, China}
\affiliation{Shenzhen Institute for Quantum Science and Engineering and Department of Physics, Southern University of Science and Technology, Shenzhen 518055, China}
\affiliation{Guangdong Provincial Key Laboratory of Quantum Science and Engineering,
Southern University of Science and Technology, Shenzhen 518055, China}
\affiliation{Shenzhen Key Laboratory of Quantum Science and Engineering, Southern University of Science and Technology, Shenzhen,518055, China}

\author{Tao Xin}
\email{xint@sustech.edu.cn}
\affiliation{International Quantum Academy, Shenzhen 518048, China}
\affiliation{Shenzhen Institute for Quantum Science and Engineering and Department of Physics, Southern University of Science and Technology, Shenzhen 518055, China}
\affiliation{Guangdong Provincial Key Laboratory of Quantum Science and Engineering,
Southern University of Science and Technology, Shenzhen 518055, China}
\affiliation{Shenzhen Key Laboratory of Quantum Science and Engineering, Southern University of Science and Technology, Shenzhen,518055, China}

\author{Dapeng Yu}
\email{yudp@sustech.edu.cn}
\affiliation{International Quantum Academy, Shenzhen 518048, China}
\affiliation{Shenzhen Institute for Quantum Science and Engineering and Department of Physics, Southern University of Science and Technology, Shenzhen 518055, China}
\affiliation{Guangdong Provincial Key Laboratory of Quantum Science and Engineering,
Southern University of Science and Technology, Shenzhen 518055, China}
\affiliation{Shenzhen Key Laboratory of Quantum Science and Engineering, Southern University of Science and Technology, Shenzhen,518055, China}

% \maketitle

% %%%%%%%%%%%%%%%%%%%%%%%%%%%%%
% \appendix
% %%%%%%%%%%%%%%%%%%%%%%%%%%%%%

\date{\today}

\maketitle

% %%%%%%%%%% Prefix a "S" to all equations, figures, tables and reset the counter %%%%%%%%%%
% \setcounter{equation}{0}
% \setcounter{figure}{0}
% \setcounter{table}{0}
% %\setcounter{page}{1}

% \renewcommand{\theequation}{S\arabic{equation}}
% \renewcommand{\thefigure}{S\arabic{figure}}
% \renewcommand{\bibnumfmt}[1]{[S#1]}
% \renewcommand{\citenumfont}[1]{S#1}

%\newpage
%\begin{center}
%\tableofcontents 
%\end{center}

\onecolumngrid

\tableofcontents
\section{DEVICE AND EXPERIMENTAL SETUP}
\subsection{Device and wiring}

\begin{figure}[H]
	\centering
	\includegraphics[width=100mm]{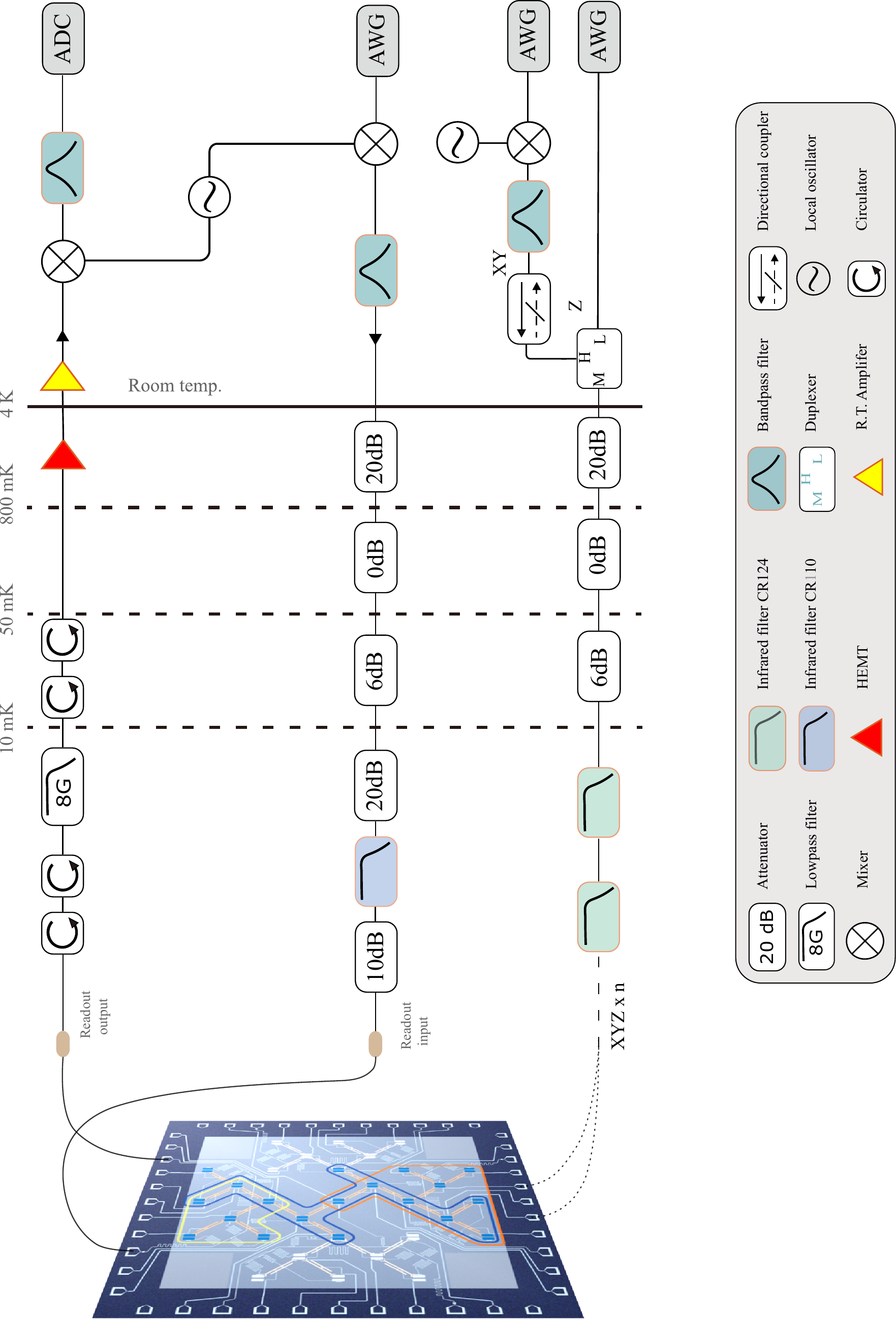}
	\caption{
		\textbf{Electronics and quantum processor schematic diagram of the experimental setup}. (Top) The schematic diagram of room temperature control electronics and wiring. (Bottom)  The schematic diagram illustrates the tree-like superconducting qubit processor comprising four transmission lines, their associated cavities, and corresponding qubits. The qubits employed in our experiments are depicted in blue. For the 6-, 9-, and 12-qubit experiments, the qubits were circled in yellow, orange, and blue, respectively.}
	\label{fig: wiring}
\end{figure}

We conducted the experiments with a superconducting processor to validate our conclusions. Figure~\ref{fig: wiring} illustrates the superconducting qubit chip employed in this work. The chip utilizes a flip-chip package featuring a bottom layer with control lines for qubits, magnetic flux control lines for couplers, readout cavities, and a Purcell filter. Notably, we combine the qubits' control lines and neighboring couplers' flux lines. 
The top layer consists of fixed-frequency qubits connected in a binary tree topology via tunable-frequency couplers. By dynamically tuning coupler frequencies using flux pulses, we enable the implementation of CZ gates between neighboring qubits.
Adjacent qubits are designed with alternating high and low frequencies of around 4.63 GHz and 4.15 GHz in average with an anharmonicity of approximately 0.20 GHz. Tunable couplers are designed with a maximum frequency of around 5.60 GHz, and the readout cavity frequency is about 6.52 GHz.

We tried our best to minimize the crosstalk of XY and Z signals in our chip. On the one hand, we maximized the separation between Co-Planar Waveguide (CPW) transmission lines.  
On the other hand, we introduced a process to cover the control lines with an additional metal layer, creating a coverage bridge across the control lines.
This method confines signals locally, preventing long-range crosstalk. For the ZZ crosstalk, it's impractical to eliminate ZZ coupling between qubits. However, our design with optimized parameters allowed us to mitigate ZZ coupling by idling the coupler frequency at specific positions. These considerations enabled us to achieve high-fidelity parallel single-qubit gates. Finally, the chip was cooled to a base temperature of 10 mK in a dilution refrigerator. 

%Fig-S1 - Device and electronics

To effectively control single qubits, we require a programmable microwave pulse signal that resonates with the qubit. However, we cannot directly generate such high-frequency arbitrary waveforms due to the limitations of arbitrary waveform generators (AWG) sampling rates in our setup. Initially, we create a pulse signal of approximately 600 MHz using an AWG. Subsequently, we mixed the signal with a 4.0 GHz Local Oscillator (LO) signal to generate the necessary single-qubit control pulses. However, the mixed signal still contains leaked LO signals and mirrored signals. We employ a bandpass filter to eliminate unwanted signals. The implementation of CZ gate signals is relatively straightforward and can be directly generated by the AWG. 
The XY and Z signals are combined with duplexers which are then fed into the processor's multiplexed control lines.
%After being combined by a duplexer, the Z and XY signals are added to the processor's multiplexed control lines.  Following filtration through
The generation of the readout signals is similar to that of single-qubit control signals. After Purcell filters, the readout signals is amplified using HEMTs and room temperature amplifiers before down-conversion and then digitized by an analog-to-digital converter (ADC). 

\subsection{Device parameters}

In our twelve-qubit experiment, we utilized qubits with parameters detailed in Table \ref{table:Device}. Meanwhile, we conducted the experiments with another chip involving six and nine qubits, sharing identical design parameters. Here, we avoid reiterating the similar parameters on the second chip. Among the selected twelve qubits, we achieved the average $T_{1}$ relaxation time of 55 $\mu s$, and the average $T_{2}$ coherence time of 24 $\mu s$. Concerning readout fidelities for the qubits, we achieved the median fidelity of 96.15\% for reading the $\ket{0}$ state and 91.45\% for the $\ket{1}$ state.

Furthermore, we employed cross-entropy benchmarking (XEB) to calibrate the fidelity of our single-qubit and two-qubit gates. The median fidelity of simultaneously operated single-qubit gates reached 99.92\%, while individually operated two-qubit gates achieved a fidelity of 97.80\%. 

\begin{table}[htp]

	\renewcommand
	\arraystretch{2.5}
	\centering

	\begin{tabular}{|p{4cm}<{\centering}|p{1cm}<{\centering}|p{1cm}<{\centering}|p{1cm}<{\centering}|p{1cm}<{\centering}|p{1cm}<{\centering}|p{1cm}<{\centering}|p{1cm}<{\centering}|p{1cm}<{\centering}|p{1cm}<{\centering}|p{1cm}<{\centering}|p{1cm}<{\centering}|p{1cm}<{\centering}|}

		\hline
		\hline

		Qubit$^{\rm a}$                           & $\text{Q}_1$  & $\text{Q}_2$  & $\text{Q}_3$  & $\text{Q}_4$  & $\text{Q}_5$  & $\text{Q}_6$  & $\text{Q}_7$  & $\text{Q}_8$  & $\text{Q}_9$   & $\text{Q}_{10}$ & $\text{Q}_{11}$ & $\text{Q}_{12}$ \\
		\hline
		Frequency (GHz)                           & 4.189  & 4.626  & 4.180  & 4.599  & 4.109  & 4.128  & 4.659  & 4.208  & 4.148   & 4.576    & 4.072    & 4.160    \\
		Anharmonicity (MHz)                       & -187   & -177   & -189   & -175   & -185   & -217   & -177   & -190   & -187    & -175     & -182     & -180     \\
		Resonator frequency (GHz)                 & 6.336  & 6.656  & 6.422  & 6.580  & 6.411  & 6.473  & 6.316  & 6.451  & 6.438   & 6.634    & 6.534    & 6.367    \\
		Resonator linewidth (MHz)                 & 2.28   & 1.12   & 1.22   & 1.70   & 1.11   & 4.94   & 0.43   & 2.91   & 1.25    & 2.67     & 3.13     & 3.14     \\
		Dispersive shift of $|1\rangle$ (MHz)     & 0.75   & 1.34   & 1.00   & 1.30   & 1.34   & 1.33   & 1.87   & 1.33   & 1.33    & 1.33     & 1.07     & 1.33     \\
		Readout fidelity of $|0\rangle$ (\%)      & 97.0   & 98.0   & 97.0   & 95.3   & 95.2   & 94.6   & 95.0   & 97.5   & 97.1    & 98.0     & 91.9     & 93.5     \\
		Readout fidelity of $|1\rangle$ (\%)      & 95.0   & 93.4   & 93.7   & 80.2   & 92.6   & 90.0   & 91.7   & 90.3   & 91.2    & 94.4     & 88.1     & 90.9     \\
		Relaxation time of $|1\rangle$  ($\mu s$) & 77.1   & 57.2   & 54.0   & 43.1   & 73.8   & 53.4   & 39.7   & 22.7   & 47.3    & 69.9     & 72.9     & 54.1     \\
		Ramsey decay time ($\mu s$)               & 39.7   & 24.4   & 12.5   & 17.9   & 15.9   & 21.6   & 15.3   & 12.9   & 21.7    & 27.0     & 24.2     & 60.4     \\
		Spin echo decay time ($\mu s$)            & 25.6   & 82.1   & 20.4   & 107.0  & 42.3   & 25.8   & 26.8   & 43.3   & 10.6    & 49.3     & 94.1     & 29.8     \\
		%1-Q gate error$^{\rm b}$ (simul.)(\%)           & 0.08  & 0.04  & 0.21  & 0.18  & 0.07  & 0.13  & 0.12  & 0.96  & 0.15  & 0.04  & 0.03  & 0.04  \\
		1-Q gate error$^{\rm b}$ (simul.)(\%)     & 0.06   & 0.03   & 0.16   & 0.14   & 0.05   & 0.10   & 0.09   & 0.72   & 0.11    & 0.03     & 0.02     & 0.03     \\

		\hline
		\hline
		CZ gate                                   & $C1-2$ & $C2-3$ & $C3-4$ & $C4-5$ & $C5-6$ & $C6-7$ & $C7-8$ & $C7-9$ & $C9-10$ & $C10-11$ & $C10-12$ & $-$      \\
		\hline
		%CZ gate error$^{\rm c}$ (\%)         &  2.34  &  2.13  &  3.21  &  7.41  &  2.91  &  2.18  & 3.69  &  2.62  &  1.54   &  1.17    &  1.42    &     - \\
		CZ gate error$^{\rm c}$ (\%)              & 2.19   & 2.00   & 3.01   & 6.95   & 2.73   & 2.04   & 3.46   & 2.46   & 1.44    & 1.10     & 1.33     & -        \\
		\hline
		\hline
	\end{tabular}

	\begin{tablenotes}
		\footnotesize

		\item [a] a. The parameters presented in the table are measured while the couplers are idle at the near ZZ coupling closed point. They represent a snapshot of our experimental process. 
  
  b. We simultaneously employ cross-entropy benchmarking (XEB) to estimate the fidelity of single-qubit gates and ascertain the Pauli error based on the decay rate. 
  
  c. The CZ gates Pauli errors are also assessed by XEB.

	\end{tablenotes}
	\caption{\textbf{Device parameters}.}
	\label{table:Device}
\end{table}

\subsection{Readout crosstalk and correction}

%State preparation and measurement errors result in inaccuracies during qubit measurements, introducing additional bias to the desired expected value. An approach to mitigate the readout error is described as follows:

 State preparation and measurement (SPAM) errors, coupled with intrinsic quantum circuit noise, can significantly degrade the fidelity of the experimental outcomes. To mitigate the impact of SPAM errors, especially under conditions of minimal initialization errors, we can model measurement errors using the following equation:
\begin{equation}
	\begin{aligned}
		\xi_{\text{noisy}} = E \xi_{\text{ideal}},
	\end{aligned}
\end{equation}
where $\xi_{ideal}$ is the vector of ideal qubit populations, $\xi_{\text{noisy}}$ is the vector of measurement qubit populations, $E$ is the response matrix, describing the transition from an ideal probability distribution to noisy distribution, the size of $E$ scales exponentially with the number of qubits $N$.

% For the distribution of the original measurement probability, the inclusion of zero values in the computational basis does not impact the calculation of $\xi_{ideal}$, {\color{red}while can increase computational resources}.

\begin{figure}[H]
	\centering
	\includegraphics[width=150mm]{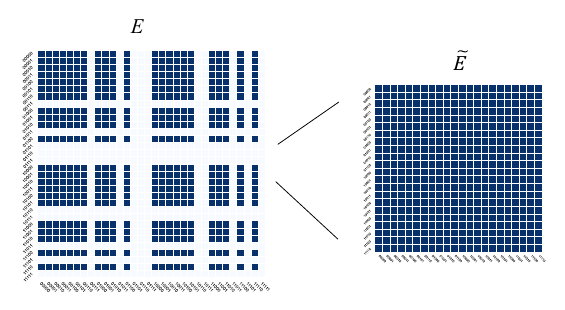}
	\caption{
		\textbf{Example for modified response matrix}. Matrix $E$ is the element assignment matrix corresponding to the non-zero term distribution of the computational basis for the single shot outcome in a 5-qubit experiment. In this matrix, blue elements represent non-zero terms, while white elements represent zero terms. $\widetilde{E}$ is constructed based on the corresponding none-zero terms and re-normalized by columns.}
	\label{fig: MThree}
\end{figure}

We illustrate the readout mitigation approach called Mthree \cite{nation2021scalable} for the 12-qubits case. Assuming minor readout crosstalk, the response matrix $E$ can be constructed by taking the tensor product of single qubit readout matrices. For the distribution of the original measurement probability, the inclusion of zero values in the computational basis does not impact the calculation of $\xi_{\text{ideal}}$, while it can increase computational resources. Mthree performs error mitigation by re-normalizing the response matrix $E$ to obtain a modified response matrix $\widetilde{E}$, which only consists of rows and columns corresponding to computational basis with non-zero counts. As illustrated in Fig.~\ref{fig: MThree}, error mitigation is performed using the following equation:

\begin{equation}
	\begin{aligned}
		\widetilde{\xi}_{\text{ideal}} = \widetilde{E}^{-1} \widetilde{\xi}_{\text{noisy}},
	\end{aligned}
	\label{eq: lux}
\end{equation}
%The matrix $\widetilde{A}$ is subjected to a lower-upper ($LU$) decomposition to obtain an upper and lower triangular matrix. 
$\widetilde{E}$ is an assignment matrix restricted to the bit strings observed in ${\xi}_{\text{noisy}}$. The columns of $\widetilde{E}$ are renormalized to ensure they sum to one.

${\xi}_{\text{noisy}}$ represents the probability distribution obtained from experiments.
We only consider the non-zero elements to get $\widetilde{\xi}_{\text{noisy}}$, which is used for readout mitigation. 
By solving the Eq. \eqref{eq: lux}, we get $\widetilde{\xi}_{\text{ideal}}$. However, it is possible for $\widetilde{\xi}_{\text{ideal}}$ to contain negative values. In such cases, Mthree provides maximum likelihood estimation (MLE) to transform the probability distribution into a non-negative form, ensuring that the resulting probability distribution is non-negative and normalized.

%Fig S2 - Multiqubit crostalk

\section{THE PREPARATION CIRCUIT OF STATES} 
In this section, we give the specific details of designing quantum circuits to realize the quantum states presented in the main text.
We have selected $N$-qubit W states, the ground state of full-connected Hamiltonian, and random circuit states as examples to illustrate our work. Next, we introduce these preparation circuits in detail.

\subsection{Design of W state preparation circuit}
The W state, named after Wolfgang D{\"u}r \cite{dur2000three}, represents multipartite entanglement as an equal-weight superposition of all terms with one qubit in $\ket{1}$ and all others in $\ket{0}$.  Together with the common GHZ (Greenberger–Horne–Zeilinger) state \cite{eltschka2012entanglement,song201710}, they represent two different kinds of multipartite entanglement. The W state has wide applications in quantum communication \cite{joo2003quantum}, quantum computing \cite{walther2005experimental}, quantum memory \cite{choi2010entanglement}, and quantum sensing \cite{li2023quantum}. In this part, we design a circuit scheme for W state preparation which is not limited by qubit number and qubit connection structure and still convenient to be implemented.  An $N$-qubit W state is expressed as,

\begin{equation}
	|W_N\rangle= \frac{1}{\sqrt{N}}(\underbrace{|100 \ldots 0\rangle}_{N- \text{qubit\;state}} + |010 \ldots 0\rangle+ \cdots+|000 \ldots 1\rangle).
	\label{eq:N_wst}
\end{equation}

 We design a recursive method for generating a preparation circuit for $N$-qubit W state with a general number $N$, which can work in universal superconducting qubit chip structures, requiring only the value of $N$ and the tree-like structure of how these qubits are connected. We start by introducing a special 2-qubit composite block gate denoted as B($p$). The B($p$) gate performs a specific function that can be summarized as follows \cite{yesilyurt2015optical,yesilyurt2016deterministic,cruz2019efficient}:

\begin{equation}
	\text{B}(p)=\left(\begin{array}{cccc}
			1 & \quad 0 & 0          & 0           \\
			0 & \quad 0 & \sqrt{1-p} & \sqrt{p}    \\
			0 & \quad 0 & \sqrt{p}   & -\sqrt{1-p} \\
			0 & \quad 1 & 0          & 0
		\end{array}\right),\quad
	\begin{aligned}
		 & \text{B}(p)|00\rangle=|00\rangle,                              \\
		 & \text{B}(p)|10\rangle=\sqrt{p}|10\rangle+\sqrt{1-p}|01\rangle.
	\end{aligned}
	\label{eq:Bp_gate}
\end{equation}

\begin{figure}[H]
	\centering
	\includegraphics[width=160mm]{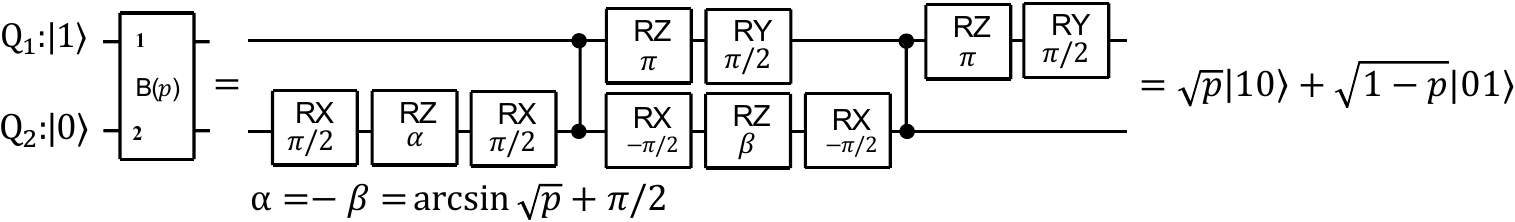}
	\caption{
		A decomposition of the B($p$) gate into a finite sequence of elementary gates on a superconducting quantum processor. RX, RY, and RZ are single-qubit rotations around $x, y$, and $z$ axes with the angle in the bottom, respectively. The lines across two qubits denote Controlled-Z (CZ) gates.}
	\label{fig: bpgate}
\end{figure}

The circuit of B($p$) gate is presented in Fig.~\ref{fig: bpgate}, where the input state of qubit $\text{Q}_1$ is $\ket{1}$, and the input state of qubit $\text{Q}_2$ is $\ket{0}$.

B($p$) gate redistributes the weight between $|10\rangle$ and $|01\rangle$ based on the parameter `$p$' when input is $|10\rangle$ while leaving $|00\rangle$ unchanged, which creates the W state by averaging the weight of $|1\rangle$ from one qubit to all $N$ qubits equally. To achieve this target in arbitrary qubit chip structures, the parameters $\{p_i\}$ should be well designed. We present the pseudocode for obtaining preparetion circuit of W states in Table~\ref{code_wst_cir}.

\begin{table}[H]
	\centering
	\begin{tabular}{l}
		\hline\hline
		\textbf{Procedure:} W state preparation circuit deduction in specified structure.               \\[0.5pt]
		\hline
		\textbf{Input}: Tree-like structure with qubit number $N$ and $i=0$      \\[0.5pt]
		\textbf{Output}: W state preparation circuit with parameters ${p_{i}}$ in each $\text{B}(p_{i})$ gate  \\[0.5pt]
		0. Initial setting: select a qubit as the first node $n_{0}$ and implement a NOT gate on it      \\
		1. While $i<N-1$ do                                                                         \\
		\quad\quad if node has branches $b_{i}>0$ do                                                     \\
		\quad\quad\quad select a branch and denote the node on other side of branch as $n_{i+1}$ \\
  
  \quad\quad\quad count nodes number $c_{i}$ behind the node $n_{i}$ (include itself) and $c_{i+1}$ for the node $n_{i+1}$  \\
		\quad\quad\quad calculate the parameter $p_{i+1} \leftarrow (c_{i}-c_{i+1})/c_{i}$                         \\
		\quad\quad\quad implement block gate $\text{B}(p_{i+1})$ on the qubits denoted by $n_i$ and $n_{i+1}$        \\
        \quad\quad\quad update the parameter $b_{i} \leftarrow b_{i}-1,\;c_{i}  \leftarrow c_{i}-c_{i+1}$                                \\
		\quad\quad\quad update to next node $i \leftarrow i+1$                                                     \\
		\quad\quad else $b_{i}=0$ do                                                                    \\
		\quad\quad\quad trace back through the branch and assign the last index to the node on the other side \\
		\quad\quad end if                                                                               \\
		\quad End while and output circuit and parameters $\{p_{i}\}$                                   \\
		\hline\hline
	\end{tabular}
	\caption{\textbf{Pseudocode for designing W state preparation circuit.}}
	\label{code_wst_cir}
\end{table}

\begin{figure}[H]
	\centering
	\includegraphics[width=150mm]{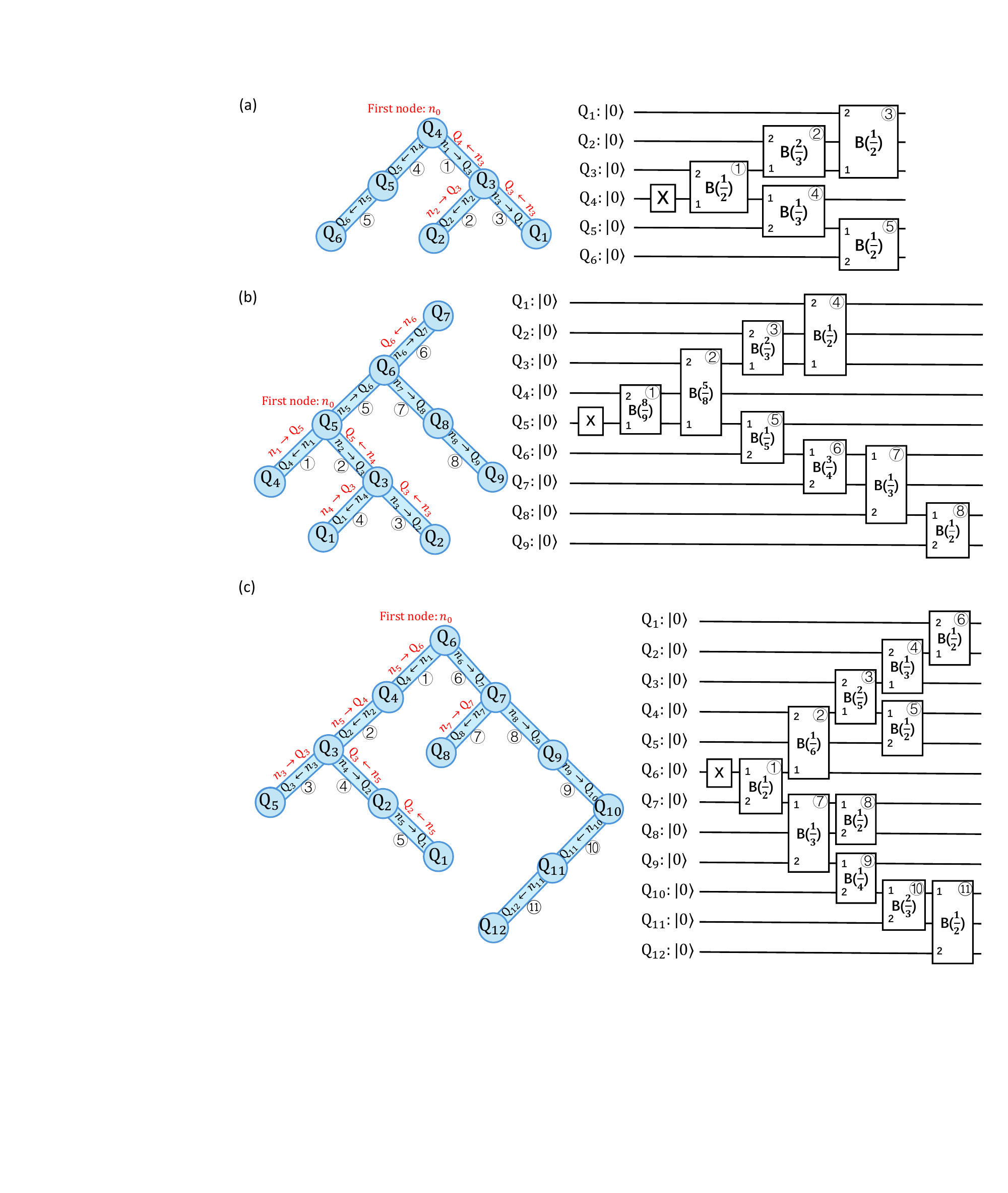}
	\caption{
		\textbf{The illustration of 6-, 9-, 12-qubit W state preparation circuits.} Left side of (a)  describes the deduction of parameters $\{p_i\}$ in a specified 6-qubit connection structure. The nodes are denoted by $\{n_i\}$ and $\text{B}(p_i)$ gates are represented by branches with black arrows on it, and the numbers with a circle below the branches represent the deducing order of $\text{B}(p_i)$ gates. Red arrows depict the reassignments of the current index $i$ to the node on the other side through the pruned branch. The right side of (a) is the modular circuit for preparing 6-qubit W state. (b) and (c) illustrate the deductions of 9- and 12-qubit W state preparation circuits and the modular circuit results, respectively.}
	\label{fig_wst_6q_cir}
\end{figure}

The process in Table~\ref{code_wst_cir} contains three kinds of operations as shown in Fig.~\ref{fig_wst_6q_cir}. (1) Update the current index $i$ and assign the updated notation $n_{i+1}$ to one of the nodes behind the current $n_i$ through a branch. Here, each node denotes a qubit and each branch denotes a block gate $\text{B}(p_i)$. (2) Deduce the value of the parameter $p_{i+1}$ of the $(i+1)$-th block gate $\text{B}(p_{i+1})$ between the current node $n_i$ and the updated node $n_{i+1}$. (3) If there is no branch behind the node $n_{i+1}$, then reverse through the pruned branch and reassign the notation $n_{i+1}$ to the node on the other side.

We illustrate this process by taking a 6-qubit W state preparation circuit as an example. According to the tree-structure figure in the left side of Fig.~\ref{fig_wst_6q_cir}(a), we start from $\text{Q}_4$ as the first node with red notation $n_0$ above it. \\
\textbf{First step:} Node $n_0$ has $b_0=2$ branches behind it, and we select the branch between $\text{Q}_4$ and $\text{Q}_3$ as the first block gate $\text{B}(p_{1})$, then the notation of $\text{Q}_3$ updates to $n_1$ as a black arrow in the branch \ding{172}, and $b_0$ changes to 1. $n_0$ has $c_0=6$ nodes behind it (include itself) and $n_1$ has $c_1=3$ nodes behind it, thus $p_{1}=(6-3)/6=1/2$. \\
\textbf{Second step:} The node $n_1$ which has two branches connecting to $\text{Q}_2$ and $\text{Q}_1$. We choose $\text{Q}_2$ as node $n_2$ with updated index $i=2$ and then $p_2$ is $(3-1)/3=2/3$ because there is only $c_2=1$ node behind this branch. Because the branch number $b_{2}$ of node $n_2$ is zero and there is no branch after it, the index is not updated. In this case, it should reverse back to the previous node $n_1$, and the corresponding qubit $\text{Q}_3$ will then be reassigned to $n_2$, as presented by the red arrow above the branch \ding{173}. \\
Similarly, the following steps shown in the branches \ding{174}, \ding{175}, and \ding{176} can be performed. Upon completing the procedure outlined above, the preparation circuit of the 6-qubit W state is shown on the right side of Fig.~\ref{fig_wst_6q_cir}(a). We also plot the design process of the state preparation circuits of 9-, 12-qubit W states in Fig.~\ref{fig_wst_6q_cir}(b-c).

\begin{figure}[H]
	\centering
	\subfloat[6-qubit quantum circuit for preparing the ground state of $\mathcal{H}_{\mathrm{FC}}$.]{
		\includegraphics[width=120mm]{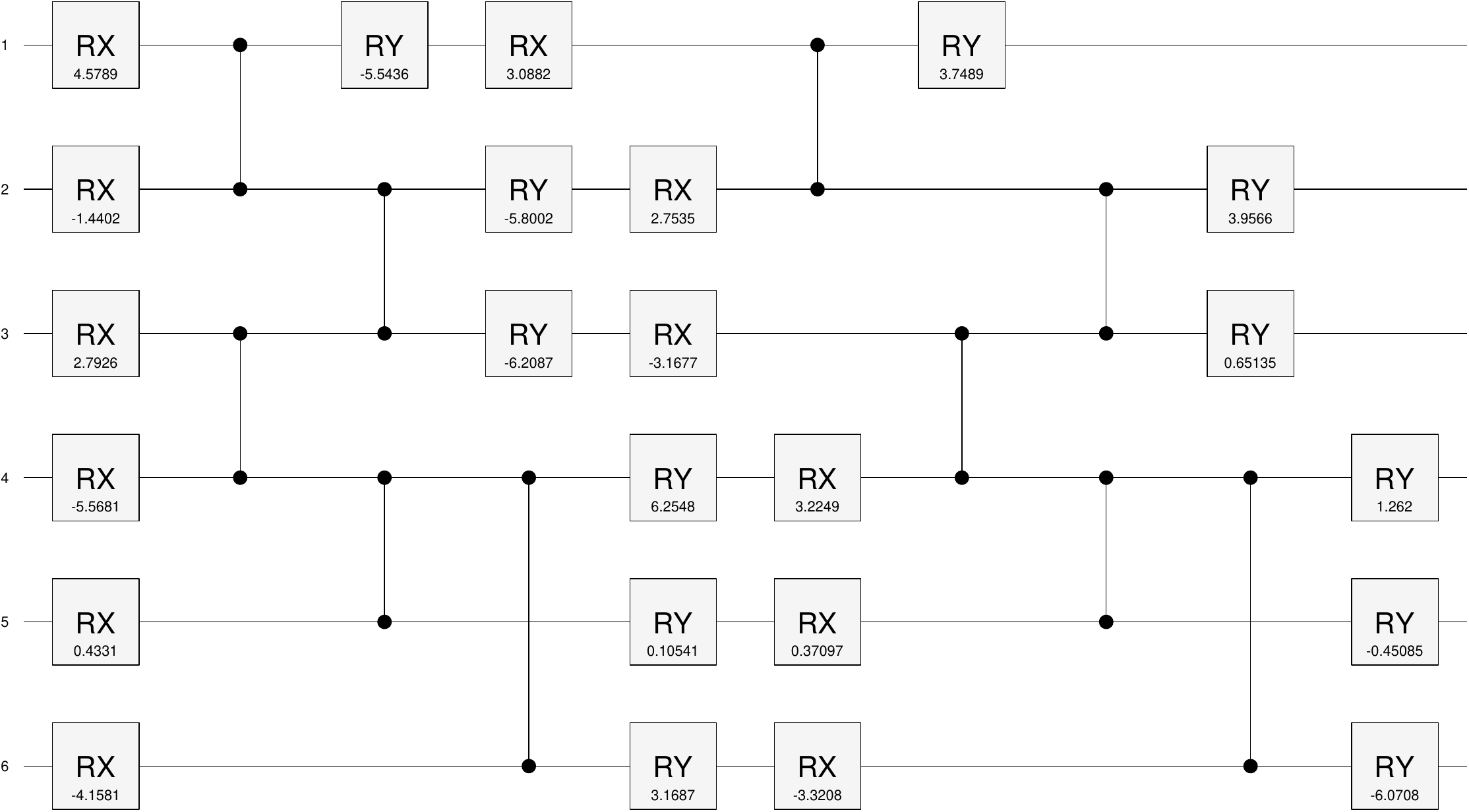}
		\label{gst_6q_cir}
	}
 
	\subfloat[6-qubit quantum circuit for preparing the random state.]{
		\includegraphics[width=120mm]{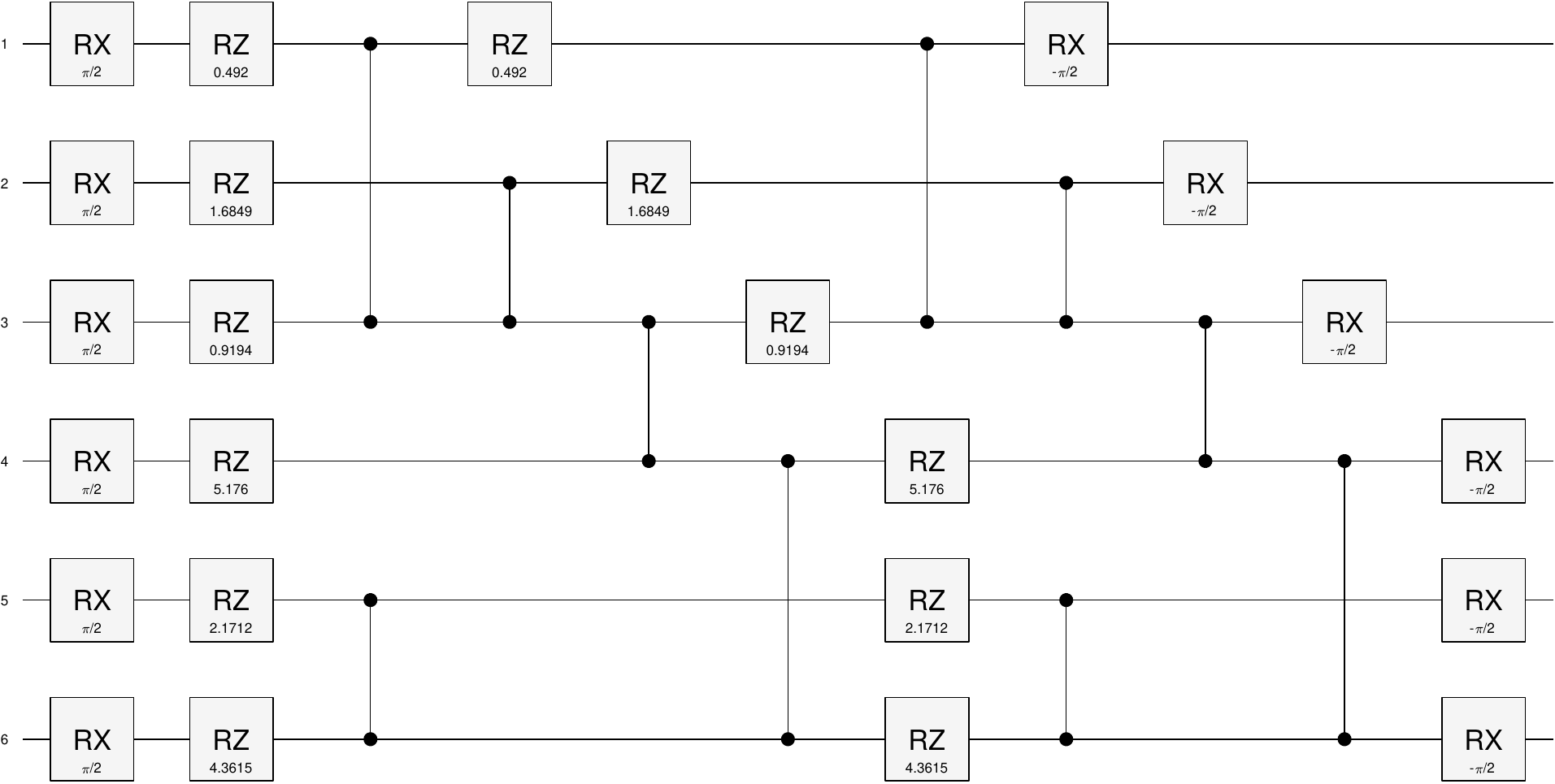}
		\label{rst_6q_cir}
	}

%	\subfloat[9-qubit random circuit state]{
%		\includegraphics[width=120mm]{Figs/rst_9q_cir.pdf}
%		\label{rst_9q_cir}
%	}
	\caption{\textbf{Experimental quantum circuits for preparing different states.}}
	\label{fig:exp_st_cir}
\end{figure}

\subsection{Design of the ground state and random state preparation circuits}

To demonstrate the feasibility of our method to different types of states, we also prepare the ground states of 6-qubit fully-connected (FC) Hamiltonians and random states, and then we reconstruct their density matrices using different tomography methods. Here, we consider the following coupling Hamiltonian model,
\begin{equation}
	 \mathcal{H}_{\mathrm{FC}}=\sum_{i=1, j>i}^{i=6} \sum_{\alpha, \beta=0}^{3} J_{\alpha, \beta}^{(i, j)} \sigma_\alpha^{(i)} \sigma_\beta^{(j)}+\sum_{k=1}^6 \sum_{l=0}^{3} \omega_l^{(k)} \sigma_l^{(k)}.
\label{eq:HFC}
\end{equation}
This Hamiltonian is comprised of all two-qubit coupling terms and single-qubit terms, with corresponding parameters $\{J_{\alpha, \beta}^{(i, j)}\}$ and $\{\omega_l^{(k)}\}$. $\sigma_0=\mathbbm{1}$ and $\sigma_{1,2,3}=\sigma_{x,y,z}$ represent the Pauli matrices. To prepare the ground state of this Hamiltonian, we use a variational quantum eigensolver that prepares the ground state by minimizing the energy $\langle\mathcal{H}_{\mathrm{FC}}\rangle=\bra{\psi(\bm{\theta})}\mathcal{H}_{\mathrm{FC}}\ket{\psi(\bm{\theta})}$ \cite{tilly2022variational}. ${\psi(\bm{\theta})}=U(\bm{\theta}^{(2)})U(\bm{\theta}^{(1)})\ket{0}^{\otimes 6}$ is the parameterized quantum state. Considering the qubit connection structure of our superconducting chip, as shown in Fig.~\ref{fig:exp_st_cir}(a), the $l$-th layer parameterized circuit evolution is adopted as,
\begin{equation}
	 U(\bm{\theta}^{(l)})=\prod _{i=1}^6 \text{RY}_i(\theta_i^{(l)})\cdot \text{CZ}_{46}\text{CZ}_{45}\text{CZ}_{23}\text{CZ}_{34}\text{CZ}_{12}\cdot\prod _{i=1}^6 \text{RX}_i(\theta_{i+6}^{(l)}).
\end{equation}
We optimized the circuit parameters $\bm{\theta}$ to reach the ground state with a numerical fidelity of over 98\%. For random states, we randomly choose the circuit which covered all possible qubit connections according to the superconducting qubit chip structure by CZ gates with random generated single-qubit rotaion gates, as shown in Fig.~\ref{fig:exp_st_cir}(b). After that, we reconstruct their density matrices using FQST, LQST, and PQST, respectively.

\section{QUANTUM STATE MEASUREMENT AND RECONSTRUCTION}

\subsection{Principle of quantum overlapping tomography}

Given a quantum state $\rho$ in an $N$-qubit quantum system, it can be expanded as,

\begin{equation}
\rho=\frac{1}{2^N}\sum_{i,j,...,l=0}^3\pi_{ij...l}\underbrace{\sigma^{(1)}_i\otimes \sigma^{(2)}_j ... \otimes \sigma^{(N)}_l}_{N\;\text{qubits}},
\end{equation}
where $\sigma_0=\mathbbm{1}$ and $\sigma_{1,2,3}=\sigma_{x,y,z}$ are the Pauli matrices. $\{\pi_{i j \ldots l}\}$ represents the expansion parameter associated with the corresponding Pauli observable $\sigma^{(1)}_i\otimes \sigma^{(2)}_j  ... \otimes \sigma^{(N)}_l$. FQST requires performing quantum measurements in the complete Pauli observables set to reconstruct the density matrix. LQST measures the expectation values of local Pauli observables involved in all $k$-qubit reduced density matrices (RDMs). This requires measuring $D_{\text{LQST}}=3^k {N\choose k}$ local Pauli observables. However, by exploring parallel measurements, the number of required measurement observables can be substantially reduced.
For instance, consider measuring a $3$-qubit operator $\sigma^{(1)}_{z}\otimes\sigma^{(2)}_{z}\otimes\sigma^{(3)}_{z}$. The expectation value of this operator is:
\begin{equation}
	\text{tr}\{(\sigma^{(1)}_{z}\otimes\sigma^{(2)}_{z}\otimes\sigma^{(3)}_{z})\rho\}=\frac{1}{2^3}(p_{\uparrow\uparrow\uparrow}-p_{\uparrow\uparrow\downarrow}-p_{\uparrow\downarrow\uparrow}+p_{\uparrow\downarrow\downarrow}-p_{\downarrow\uparrow\uparrow}+p_{\downarrow\uparrow\downarrow}+p_{\downarrow\downarrow\uparrow}-p_{\downarrow\downarrow\downarrow}),
\end{equation}
where $p$ means the probability of each outcome labeled in the subscript. Then we can infer the expectation values of some 2-qubit operators, for example, $\sigma_z^{(1)}\otimes\sigma_z^{(2)}$, it can be written as:
\begin{equation}
	\text{tr}\{(\sigma^{(1)}_{z}\otimes\sigma^{(2)}_{z}\otimes \mathbbm{1})\rho\}=\frac{1}{2^3}(p_{\uparrow\uparrow\uparrow}+p_{\uparrow\uparrow\downarrow}-p_{\uparrow\downarrow\uparrow}-p_{\uparrow\downarrow\downarrow}-p_{\downarrow\uparrow\uparrow}-p_{\downarrow\uparrow\downarrow}+p_{\downarrow\downarrow\uparrow}+p_{\downarrow\downarrow\downarrow}).
\end{equation}

The formula above clearly shows that the expectation value of the 2-qubit Pauli operator can be obtained from the probability distribution of the 3-qubit Pauli operator by tracing out the third qubit.
%From this observation, we can infer that the distribution information of the tensor product of many-site Pauli operators fully encapsulates the details of few-site Pauli operators. 
This observation implies that the measurements on parallel Pauli observables can also provide the expectation values of some local Pauli observables.
This highlights the capability of parallel measurement, which has the potential to substantially reduce measurement resources by incorporating local measurements into parallel measurements.

The quantum overlapping tomography (QOT) technique proposed by J. Cotler and F. Wilczek \cite{cotler2020quantum} (and also independently proposed by X. Bonet-Monroig et al. \cite{bonet2020nearly})  is used to reconstruct all $k$-qubit RDMs from parallel measurements. In their protocol, each qubit can be encoded in terms of a family of perfect hash functions (FPHF). Based on how these qubits are encoded, we can select a set of parallel Pauli observables for measurement. An $(N, k)$ FPHF is a family of mappings from $N$-element set $\{1,2,...,N\}$ to $k$-element set $\{0,1,...,k-1\}$, such that at least one function in this family satisfies one-to-one mapping from qubits in every $k$-qubit subsystem to the set $\{0,\dots,k-1\}$. This property of the FPHF ensures that the qubits in each $k$-qubit subsystem can be uniquely labeled, distinguishing them from one another. This unique labeling allows for a complete coverage of all $k$-qubit local Pauli observables, thereby enabling the reconstruction of all $k$-qubit RDMs.

The $(N, k)$ FPHF can be written as the  $l\times N$ matrix with $k$ accessible values, where $l$ is the number of hash functions in this family. Here, we take (3, 2) FPHF as an example. (3, 2) FPHF is the simplest FPHF with only two hash functions $h_1$ and $h_2$, the outcome of this family of functions can be written as a $2\times3$ matrix $\left[0\;0\;1;\;0\;1\;1\right]$, where each row represents a 3-to-2 hash function mapping value table $h_i:\{1,2,3\}\rightarrow\{0,1\}$. In this case, there are only ${3\choose 2} =3$ possible two-qubit subsystems $\{1,2\}$, $\{2,3\}$, and $\{1,3\}$. Then qubits in each 2-qubit subsystem can be mapped to a one-to-one outcome $\{0,1\}$ or $\{1,0\}$ at least in one hash function. $N$-qubit parallel Pauli observables can be constructed by assigning the same Pauli measurement $\sigma_{x},\sigma_{y},$ and $\sigma_{z}$ to qubits with the same hash function. For example, the hash function $[0,0,1]$ corresponds to nine parallel Pauli observables $\{XXX,XXY,XXZ,YYX,YYY,YYZ,ZZX,ZZY,ZZZ\}$, since they must remain the same Pauli measurement for the first qubit and the second qubit. In this way, there are at most $3^k$ $N$-qubit parallel Pauli observables for each hash function. By collecting all $N$-qubit parallel observables from $(N, k)$ FPHF and eliminating repeated ones, we obtain the parallel measurement observables to reconstruct all k-qubit RDMs.

\subsection{Extension of the family of perfect hash functions for $k > 2$}

In the following section, we provide a protocol for finding $(N, k)$ FPHF. The protocol based on binary integer linear optimization can give an extended solution of $(N, k)$ FPHF. The details of this method are as follows. To find the smallest possible set of $(N, k)$ FPHF, the most straightforward idea is to optimize the set containing all hash functions. Thus the first part is constructing a binary integer relation matrix denoted by $\Lambda$. $\Lambda$ gives a brief description of the one-to-one mapping relation between candidate hash functions and $k$-qubit subsystems, and it is constructed as follows. Each row of $\Lambda$ corresponds to a candidate hash function, while each column corresponds to a target $k$-qubit subsystem. Each entry in the matrix indicates whether the hash function forms a one-to-one mapping with the corresponding $k$-qubit subsystems. If the mapping is one-to-one, the entry is assigned a value of 1; otherwise, it is assigned a value of 0.

Then a vector $\mathbf{x}$ is introduced to denote the selection of hash functions. The components of $\mathbf{x}$ can only be set to 0 or 1, where 1 represents that the selection contains the corresponding hash function while 0 means no. A possible solution can be obtained by minimizing $\lVert \mathbf{x}\lVert_{1}$ under the constrain $ \Lambda^T \mathbf{x} \geq 1$ and $ x_i=0 \text { or } 1$. Here, $\lVert \mathbf{x}\lVert_{1}=\sum_i |x_i|$ denotes the sum of absolute values of vector elements, and the condition $\Lambda^T \mathbf{x} \geq 1$ ensures that the selected set of candidate hash functions can cover all $k$-qubit subsystems. When $\lVert \mathbf{x}\lVert_{1}$ is minimized, a solution of $(N, k)$ FPHF can be extracted from candidate functions based on the positions of all the entries with a value of 1 in $\mathbf{x}$. We calculated solutions of $(N,k)$ FPHF for $N=6\sim 12$ and $k=2\sim 4$, then deduced their corresponding QOT parallel Pauli observables. The amounts of hash functions in each case are shown in Table~\ref{table:Hash_result}.  
We presented the $(9,3)$ FPHF as an example in Table~\ref{table:9_3_result}. The $(9,3)$ FPHF is a $4\times9$ matrix with only 3 different values. Each row in the matrix corresponds to at most 27 nine-qubit parallel Pauli observables. By eliminating repeated Pauli observables in each row, the total parallel measurement observables set has 99 elements. The amounts of parallel measurement observables of different $N$ and $k$ are listed in Table~\ref{table:QOT_result}. We also presented the $(6,3)$, $(9,3)$ and $(12,3)$ QOT  parallel measurement observables in Tables~\ref{table:6_3_qot_tab}, \ref{table:9_3_qot_tab}, and \ref{table:12_3_qot_tab}.

\begin{table}[h!t]
	\center
	\renewcommand
	\arraystretch{1.5}
	\caption{The amount of hash functions in different $(N,k)$ cases.}
	\begin{tabular}{p{1cm}<{\centering}p{1cm}<{\centering}p{1cm}<{\centering}p{1cm}<{\centering}p{1cm}<{\centering}p{1cm}<{\centering}p{1cm}<{\centering}p{1cm}<{\centering}p{1cm}<{\centering}p{1cm}<{\centering}}
		\hline
		\hline
		FPHF & $N$=6 & $N$=7 & $N$=8 & $N$=9 & $N$=10 & $N$=11 & $N$=12 \\
		\hline
		$k$=2  & 3   & 3   & 3   & 4   & 4    & 4    & 4    \\
		$k$=3  & 3   & 4   & 4   & 4   & 5    & 6    & 10    \\
		$k$=4  & 5   & 6   & 6   & 8   & 10   & 13   & 15  \\
		\hline
	\end{tabular}
	\label{table:Hash_result}
\end{table}

\begin{table}[h!t]
	\center
	\renewcommand
	\arraystretch{1.5}
	\caption{The $(9,3)$ family of perfect hash functions.}
	\begin{tabular}{p{1cm}<{\centering}p{1cm}<{\centering}p{1cm}<{\centering}p{1cm}<{\centering}p{1cm}<{\centering}p{1cm}<{\centering}p{1cm}<{\centering}p{1cm}<{\centering}p{1cm}<{\centering}p{1cm}<{\centering}p{1cm}<{\centering}}
		\hline 0 & 1 & 0 & 2 & 0 & 2 & 2 & 1 & 1 \\
		0        & 1 & 1 & 0 & 2 & 2 & 1 & 2 & 0 \\
		2        & 0 & 1 & 0 & 0 & 1 & 2 & 2 & 1 \\
		2        & 2 & 0 & 0 & 1 & 2 & 1 & 0 & 1 \\
		\hline
	\end{tabular}
	\label{table:9_3_result}
\end{table}

\begin{table}[h!t]
	\center
	\renewcommand
	\arraystretch{1.5}
	\caption{The amount of Pauli basis in different $(N,k)$ cases.}
	\begin{tabular}{p{1cm}<{\centering}p{1cm}<{\centering}p{1cm}<{\centering}p{1cm}<{\centering}p{1cm}<{\centering}p{1cm}<{\centering}p{1cm}<{\centering}p{1cm}<{\centering}p{1cm}<{\centering}p{1cm}<{\centering}}
		\hline
		\hline
		QOT & $N$=6 & $N$=7 & $N$=8 & $N$=9 & $N$=10 & $N$=11 & $N$=12 \\
		\hline
		$k$=2 & 21  & 21  & 21  & 27  & 27   & 27   & 27   \\
		$k$=3 & 75  & 99  & 99  & 99  & 123  & 147  & 243  \\
		$k$=4 & 315 & 453 & 453 & 609 & 777  & 999  & 1155 \\
		\hline
	\end{tabular}
	\label{table:QOT_result}
\end{table}

\begin{table}[h!t]
	\center
	\renewcommand
	\arraystretch{1.5}
	\caption{The QOT  parallel measurement observables for $N=6$ and $k=3$. For convenience, we use $X$, $Y$, and $Z$ to denote Pauli measurement operators $\sigma_x$, $\sigma_y$, and $\sigma_z$. Here, we omit the tensor symbol `$\otimes$' between qubits.} 
	\begin{tabular}{p{1.5cm}<{\centering}p{1.5cm}<{\centering}p{1.5cm}<{\centering}p{1.5cm}<{\centering}p{1.5cm}<{\centering}p{1.5cm}<{\centering}p{1.5cm}<{\centering}}
		\hline XXXXXX & XXXXYY & XXXXZZ & XXXYYX & XXXZZX & XXYXXY & XXYYXX \\
		XXYYYY & XXYZZY & XXZXXZ & XXZYYZ & XXZZXX & XXZZZZ & XYXXYX \\
		XYYXXX & XYYXYY & XYYXZZ & XYYYYX & XYZZYX & XZXXZX & XZYYZX \\
		XZZXXX & XZZXYY & XZZXZZ & XZZZZX & YXXXXY & YXXYXX & YXXYYY \\
		YXXYZZ & YXYYXY & YXZZXY & YYXXXX & YYXXYY & YYXYYX & YYXZZX \\
		YYYXXY & YYYYXX & YYYYYY & YYYYZZ & YYYZZY & YYZXXZ & YYZYYZ \\
		YYZZYY & YYZZZZ & YZXXZY & YZYYZY & YZZYXX & YZZYYY & YZZYZZ \\
		YZZZZY & ZXXXXZ & ZXXZXX & ZXXZYY & ZXXZZZ & ZXYYXZ & ZXZZXZ \\
		ZYXXYZ & ZYYYYZ & ZYYZXX & ZYYZYY & ZYYZZZ & ZYZZYZ & ZZXXXX \\
		ZZXXZZ & ZZXYYX & ZZXZZX & ZZYXXY & ZZYYYY & ZZYYZZ & ZZYZZY \\
		ZZZXXZ & ZZZYYZ & ZZZZXX & ZZZZYY & ZZZZZZ &        &  \\
		\hline
	\end{tabular}
	\label{table:6_3_qot_tab}
\end{table}

\begin{table}[h!t]
	\center
	\renewcommand
	\arraystretch{1.5}
	\caption{The QOT  parallel measurement observables for $N=9$ and $k=3$. For convenience, we use $X$, $Y$, and $Z$ to denote Pauli measurement operators $\sigma_x$, $\sigma_y$, and $\sigma_z$. Here, we omit the tensor symbol `$\otimes$' between qubits.}
	\begin{tabular}{p{2.2cm}<{\centering}p{2.2cm}<{\centering}p{2.2cm}<{\centering}p{2.2cm}<{\centering}p{2.2cm}<{\centering}p{2.2cm}<{\centering}p{2.2cm}<{\centering}p{2.2cm}<{\centering}}
		\hline XXXXXXXXX & XXXXYXYXY & XXXXYYXYX & XXXXZXZXZ & XXXXZZXZX & XXXYXYYXX & XXXZXZZXX & XXYXXYXXY \\
		XXYYXXXYX & XXYYYXYYY & XXYYZXZYZ & XXZXXZXXZ & XXZZXXXZX & XXZZYXYZY & XXZZZXZZZ & XYXXXXXYY \\
		XYXYXYYYY & XYXYYXXXX & XYXZXZZYY & XYYXXXYXX & XYYXYYYYX & XYYXZZYZX & XYYYYYXXY & XYZYYZXXZ \\
		XZXXXXXZZ & XZXYXYYZZ & XZXZXZZZZ & XZXZZXXXX & XZYZZYXXY & XZZXXXZXX & XZZXYYZYX & XZZXZZZZX \\
		XZZZZZXXZ & YXXXXXYYX & YXXYXXXXY & YXXYYYXYY & YXXYZZXZY & YXYXXYYYY & YXYXYXXXX & YXYYYYYXX \\
		YXYZYZZXX & YXZXXZYYZ & YYXXXYXXX & YYXXYYYXY & YYXXZYZXZ & YYXYYXYYX & YYYXYXXYY & YYYYXXYXY \\
		YYYYXYXYX & YYYYYYYYY & YYYYZYZYZ & YYYYZZYZY & YYYZYZZYY & YYZYYZYYZ & YYZZXYXZX & YYZZYYYZY \\
		YYZZZYZZZ & YZXZZXYYX & YZYXYXXZZ & YZYYYYYZZ & YZYZYZZZZ & YZYZZYYYY & YZZYXXZXY & YZZYYYZYY \\
		YZZYZZZZY & YZZZZZYYZ & ZXXXXXZZX & ZXXZXXXXZ & ZXXZYYXYZ & ZXXZZZXZZ & ZXYXXYZZY & ZXZXXZZZZ \\
		ZXZXZXXXX & ZXZYZYYXX & ZXZZZZZXX & ZYXYYXZZX & ZYYYYYZZY & ZYYZXXYXZ & ZYYZYYYYZ & ZYYZZZYZZ \\
		ZYZXZXXYY & ZYZYYZZZZ & ZYZYZYYYY & ZYZZZZZYY & ZZXXXZXXX & ZZXXYZYXY & ZZXXZZZXZ & ZZXZZXZZX \\
		ZZYYXZXYX & ZZYYYZYYY & ZZYYZZZYZ & ZZYZZYZZY & ZZZXZXXZZ & ZZZYZYYZZ & ZZZZXXZXZ & ZZZZXZXZX \\
		ZZZZYYZYZ & ZZZZYZYZY & ZZZZZZZZZ &           &           &           &           &          \\
		\hline
	\end{tabular}
	\label{table:9_3_qot_tab}
\end{table}

\begin{table}[H]
	\center
	\renewcommand
	\arraystretch{1.5}
	\caption{The QOT  parallel measurement observables for $N=12$ and $k=3$. For convenience, we use $X$, $Y$, and $Z$ to denote Pauli measurement operators $\sigma_x$, $\sigma_y$, and $\sigma_z$. Here, we omit the tensor symbol `$\otimes$' between qubits.}
	\begin{tabular}{p{2.9cm}<{\centering}p{2.9cm}<{\centering}p{2.9cm}<{\centering}p{2.9cm}<{\centering}p{2.9cm}<{\centering}p{2.9cm}<{\centering}}
		\hline XXXXXXXXXXXX & XXXXXXXXYYYY & XXXXXXXXZZZZ & XXXXXXXYXYYY & XXXXXXXYYYYY & XXXXXXXZXZZZ \\
		XXXXXXXZZZZZ & XXXXXXYYXXXX & XXXXXXYYXXYY & XXXXXXYYYYXY & XXXXXXYYYYYY & XXXXXXYYZZZZ \\
		XXXXXXZZXXXX & XXXXXXZZXXZZ & XXXXXXZZYYYY & XXXXXXZZZZXZ & XXXXXXZZZZZZ & XXXXXYXXXYYY \\
		XXXXXZXXXZZZ & XXXXYXXXYXXX & XXXXYYXXXXYX & XXXXYYXXYYYY & XXXXYYYXXXXY & XXXXYYYXXXYY \\
		XXXXYYYYYYYY & XXXXYYZZZZYZ & XXXXYZXXYZZZ & XXXXZXXXZXXX & XXXXZYXXZYYY & XXXXZZXXXXZX \\
		XXXXZZXXZZZZ & XXXXZZYYYYZY & XXXXZZZXXXXZ & XXXXZZZXXXZZ & XXXXZZZZZZZZ & XXXYXYYXXXXX \\
		XXXYXYYYYYYY & XXXYXYYZZZZZ & XXXYYXXXYXYY & XXXZXZZXXXXX & XXXZXZZYYYYY & XXXZXZZZZZZZ \\
		XXXZZXXXZXZZ & XXYXXXXYXYXX & XXYXXXYXYXXX & XXYXXXYYYYYY & XXYXXXYZYZZZ & XXYYXXXXXXYX \\
		XXYYXXXXYYXY & XXYYYXXYYYYY & XXYYYYYXXXYY & XXYYZZZXXXYZ & XXYZZXXYZYZZ & XXZXXXXZXZXX \\
		XXZXXXZXZXXX & XXZXXXZYZYYY & XXZXXXZZZZZZ & XXZYYXXZYZYY & XXZZXXXXXXZX & XXZZXXXXZZXZ \\
		XXZZYYYXXXZY & XXZZZXXZZZZZ & XXZZZZZXXXZZ & XYXXXXXYXXYX & XYXXXYXXXYXX & XYXXXYYYXYYY \\
		XYXXXYZZXYZZ & XYXYXXXXXYXX & XYXYYYYXXYYY & XYXYZZZXXYZZ & XYYYXXXYYYYY & XYZZXXXYZZYZ \\
		XZXXXXXZXXZX & XZXXXZXXXZXX & XZXXXZYYXZYY & XZXXXZZZXZZZ & XZXZXXXXXZXX & XZXZYYYXXZYY \\
		XZXZZZZXXZZZ & XZYYXXXZYYZY & XZZZXXXZZZZZ & YXXXYYYXXXXX & YXYXXXXYYXXX & YXYXYYYYYXYY \\
		YXYXZZZYYXZZ & YXYYYXXXYXXX & YXYYYXYYYXYY & YXYYYXZZYXZZ & YXYYYYYXYYXY & YXZZYYYXZZXZ \\
		YYXXXXXYYYXX & YYXXXYYXXXXX & YYXXYYYYXXYX & YYXXYYYYYYXY & YYXXZZZYYYXZ & YYXYYYXXXXXX \\
		YYXYYYXYXYYY & YYXYYYXZXZZZ & YYXYYYYXYXYY & YYXZZYYXZXZZ & YYYXXYYYXYXX & YYYXYXXXXXXX \\
		YYYXYXXYYYYY & YYYXYXXZZZZZ & YYYYXXXXXXXX & YYYYXXXYYYXX & YYYYXXXYYYYX & YYYYXXYYXXXX \\
		YYYYXXYYYYXY & YYYYXXZZZZXZ & YYYYXYYYXYYY & YYYYXZYYXZZZ & YYYYYXYYYXXX & YYYYYYXXXXXX \\
		YYYYYYXXXXYX & YYYYYYXXYYXX & YYYYYYXXYYYY & YYYYYYXXZZZZ & YYYYYYYXXXXX & YYYYYYYXYXXX \\
		YYYYYYYYXXXX & YYYYYYYYYYYY & YYYYYYYYZZZZ & YYYYYYYZYZZZ & YYYYYYYZZZZZ & YYYYYYZZXXXX \\
		YYYYYYZZYYYY & YYYYYYZZYYZZ & YYYYYYZZZZYZ & YYYYYYZZZZZZ & YYYYYZYYYZZZ & YYYYZXYYZXXX \\
		YYYYZYYYZYYY & YYYYZZXXXXZX & YYYYZZYYYYZY & YYYYZZYYZZZZ & YYYYZZZYYYYZ & YYYYZZZYYYZZ \\
		YYYYZZZZZZZZ & YYYZYZZXXXXX & YYYZYZZYYYYY & YYYZYZZZZZZZ & YYYZZYYYZYZZ & YYZXXYYZXZXX \\
		YYZYYYYZYZYY & YYZYYYZXZXXX & YYZYYYZYZYYY & YYZYYYZZZZZZ & YYZZXXXYYYZX & YYZZYYYYYYZY \\
		YYZZYYYYZZYZ & YYZZZYYZZZZZ & YYZZZZZYYYZZ & YZXXYYYZXXZX & YZYYYYYZYYZY & YZYYYZXXYZXX \\
		YZYYYZYYYZYY & YZYYYZZZYZZZ & YZYZXXXYYZXX & YZYZYYYYYZYY & YZYZZZZYYZZZ & YZZZYYYZZZZZ \\
		ZXXXZZZXXXXX & ZXYYZZZXYYXY & ZXZXXXXZZXXX & ZXZXYYYZZXYY & ZXZXZZZZZXZZ & ZXZZZXXXZXXX \\
		ZXZZZXYYZXYY & ZXZZZXZZZXZZ & ZXZZZZZXZZXZ & ZYXXZZZYXXYX & ZYYYZZZYYYYY & ZYZYXXXZZYXX \\
		ZYZYYYYZZYYY & ZYZYZZZZZYZZ & ZYZZZYXXZYXX & ZYZZZYYYZYYY & ZYZZZYZZZYZZ & ZYZZZZZYZZYZ \\
		ZZXXXXXZZZXX & ZZXXXZZXXXXX & ZZXXYYYZZZXY & ZZXXZZZZXXZX & ZZXXZZZZZZXZ & ZZXYYZZXYXYY \\
		ZZXZZZXXXXXX & ZZXZZZXYXYYY & ZZXZZZXZXZZZ & ZZXZZZZXZXZZ & ZZYXXZZYXYXX & ZZYYXXXZZZYX \\
		ZZYYYYYZZZYY & ZZYYYZZYYYYY & ZZYYZZZZYYZY & ZZYYZZZZZZYZ & ZZYZZZYXYXXX & ZZYZZZYYYYYY \\
		ZZYZZZYZYZZZ & ZZYZZZZYZYZZ & ZZZXXZZZXZXX & ZZZXZXXXXXXX & ZZZXZXXYYYYY & ZZZXZXXZZZZZ \\
		ZZZYYZZZYZYY & ZZZYZYYXXXXX & ZZZYZYYYYYYY & ZZZYZYYZZZZZ & ZZZZXXXXXXXX & ZZZZXXXZZZXX \\
		ZZZZXXXZZZZX & ZZZZXXYYYYXY & ZZZZXXZZXXXX & ZZZZXXZZZZXZ & ZZZZXYZZXYYY & ZZZZXZZZXZZZ \\
		ZZZZYXZZYXXX & ZZZZYYXXXXYX & ZZZZYYYYYYYY & ZZZZYYYZZZYY & ZZZZYYYZZZZY & ZZZZYYZZYYYY \\
		ZZZZYYZZZZYZ & ZZZZYZZZYZZZ & ZZZZZXZZZXXX & ZZZZZYZZZYYY & ZZZZZZXXXXXX & ZZZZZZXXXXZX \\
		ZZZZZZXXYYYY & ZZZZZZXXZZXX & ZZZZZZXXZZZZ & ZZZZZZYYXXXX & ZZZZZZYYYYYY & ZZZZZZYYYYZY \\
		ZZZZZZYYZZYY & ZZZZZZYYZZZZ & ZZZZZZZXXXXX & ZZZZZZZXZXXX & ZZZZZZZYYYYY & ZZZZZZZYZYYY \\
		ZZZZZZZZXXXX & ZZZZZZZZXXXX & ZZZZZZZZXXXX &              &              & \\
		\hline
	\end{tabular}
	\label{table:12_3_qot_tab}
\end{table}

\subsection{Learning density matrices via tensor networks}

\begin{figure}[H]
	\centering
	\includegraphics[width=150mm]{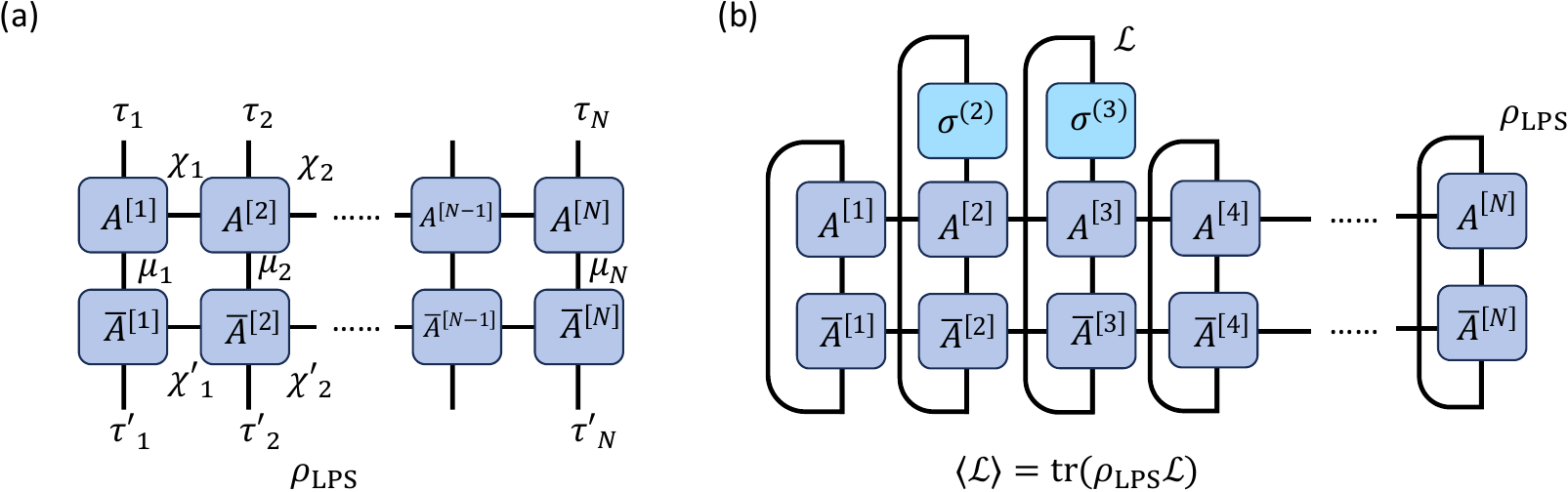}
	\caption{\textbf{The structure of LPS and its expectation value calculation}.  Its fundamental element represented by block $A^{[n]}$ is the four-bond tensor $A_{\tau_{n}, \chi_{n-1}, \chi_n, \mu_n}^{[n]}$. The factors $\chi_n$ and $\chi_{n-1}$ denote ancillary indexes, while $\tau_n$, $\mu_n$ denote the 2-dimensional physical index and purification index, repectively.}
	\label{fig:lps_tt}
\end{figure}

In this section, we introduce the quantum state tomography (QST) post-processing framework for mixed-state reconstruction.
%predicated on the concept of locally purified states (LPS). 
Here, we employ the concept of locally purified state (LPS) method to represent mixed-states \cite{glasser2019expressive}. The LPS provides a concise tensor network representation for mixed states, characterized by favorable complexity that prevents the exponential growth of the state space. As shown in Fig.~\ref{fig:lps_tt}(a), the LPS is described as a specialized form of the tensor-train. The tensors $A_{\tau_{n}, \chi_{n-1}, \chi_{n}, \mu_n}^{[n]}$ and their conjugates $\bar{A}^{[n]}$ form the LPS representation of a mixed state $\rho_{\text{LPS}}$. The physical index $\tau_{n}$ is a 2-dimension index applicable to qubits. The ancillary indexes $ \chi_{n-1}$ and $ \chi_n $ serve as the interlinking elements among distinct qubits, with their values determining the representation complexity of the state. The purification index $ \mu_n $, which connects $ A^{[n]}$ and $ \bar{A}^{[n]}$, determines the purity of the LPS state. If $\mu_n=1$, the represented state will be a pure state. The density matrix $\rho_{\text{LPS}}$ in the computational basis can be written as
\begin{equation}
	 \rho_{\text{LPS}}=\sum_{\{\boldsymbol{\tau},\boldsymbol{\tau'}\}}\sum_{\{\boldsymbol{\chi},\boldsymbol{\chi'}\}} \sum_{\{\boldsymbol{\mu}\}} \prod_{n=1}^N A^{\left[n\right]}_{\tau_{n},\chi_{n-1}, \chi_n, \mu_n}\bar{A}^{\left[n\right]}_{\tau'_{n},\chi'_{n-1}, \chi'_n, \mu_n}|\tau_{1}, \cdots, \tau_{N}\rangle\langle \tau'_{1}, \cdots, \tau'_{N}|.
	\label{eq:lps_tt}
\end{equation}
 Here, $\ket{\tau}$ is the eigenvector $\ket{0}$ or $\ket{1}$ of the Pauli matrix $\sigma_z$. The expectation values of Pauli observables can be easily computed in the LPS form. Figure ~\ref{fig:lps_tt}(b) shows the process of calculating expectation values. The local observable $\mathcal{L} = \prod_{i=1} \sigma^{(i)}$ is represented in tensor-train form, commonly referred to as a matrix product operator (MPO). The calculation of the expectation value is essentially performed by contracting the physical indexes between the LPS and the MPO. The full contraction calculation of the $\rho_{\text{LPS}}$ results in a scalar value, which is the expectation value of the observable $\mathcal{L} $ on the state $\rho_{\text{LPS}}$.

% We utilized a mean squared estimation (MSE) shceme to estimate the parameterized LPS form of the laboratory state in terms of experimental data. The MSE loss function is defined as follow,

\begin{figure}[H]
	\centering
	\includegraphics[width=180mm]{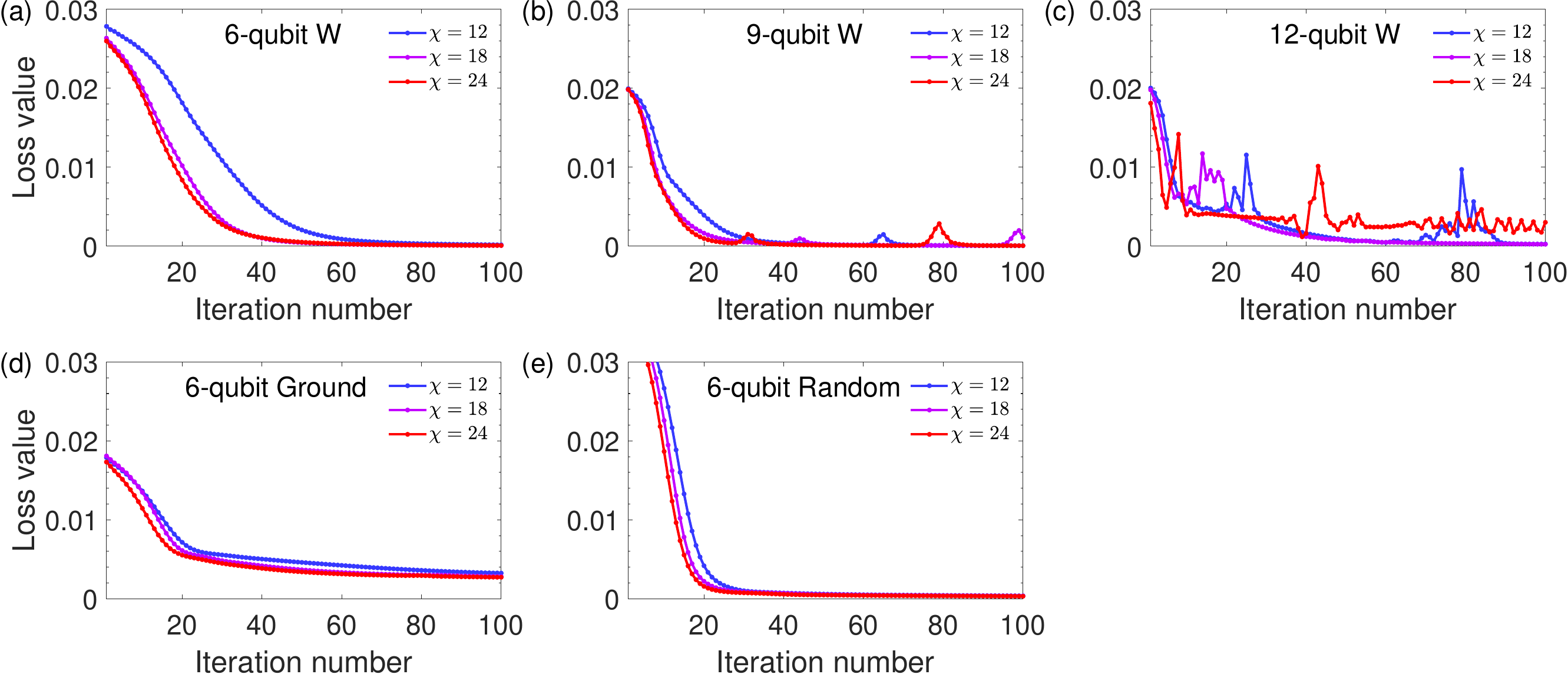}
	\caption{
		\textbf{The loss curves of learning density matrices vis LPS.} The auxiliary index dimension $\chi$ is set in [12, 18, 24], and the maximum iteration number is 100. (a), (b), and (c) are the loss value curves with the experimental measurement results of 6-, 9-, and 12-qubit W states as the inputs, respectively. (d) corresponds to the ground state of the 6-qubit FC Hamiltonian and (e) denotes the results of the 6-qubit random state.}
	\label{fig:loss_value}
\end{figure}

We employed the gradient descending algorithm in terms of the loss functions  $f_{\text{MSE}}$ and $f_{\text{MLE}}$ to approximate the parameterized LPS form of the experimentally-prepared state, utilizing experimental data. $f_{\text{MSE}}$ is a loss function defined using the Mean Squared Error (MSE) estimation , which is commonly used in many fields \cite{prasad1990estimation,cai2016optimal}. $f_{\text{MLE}}$ is another loss function defined using the Maximum Likelihood Estimation (MLE) that searches model parameters for an assumed probability distribution \cite{paris2004quantum,shang2017superfast}. The definitions of these loss functions are provided in the main text. The process of the gradient descent iterations is depicted in Fig.~\ref{fig:loss_value}. To calibrate the parameters of the LPS method, taking the minimization of $f_{\mathrm{MSE}} $ as an example, we present the LPS learning process of PQST ($k=3$)  for 6-, 9-, 12-qubit W states and the 6-qubit ground state of the FC Hamiltonian in Eq. (\ref{eq:HFC}) and the 6-qubit random state, examining the influence of the auxiliary index dimension $\chi$ on the results. It indicates that the loss values nearly reach a plateau after around 60 iterations, and the index dimension $\chi=18$ is chosen.

\subsection{The applications of PQST on dynamical states}

\begin{figure}[H]
	\centering
	\includegraphics[width=180mm]{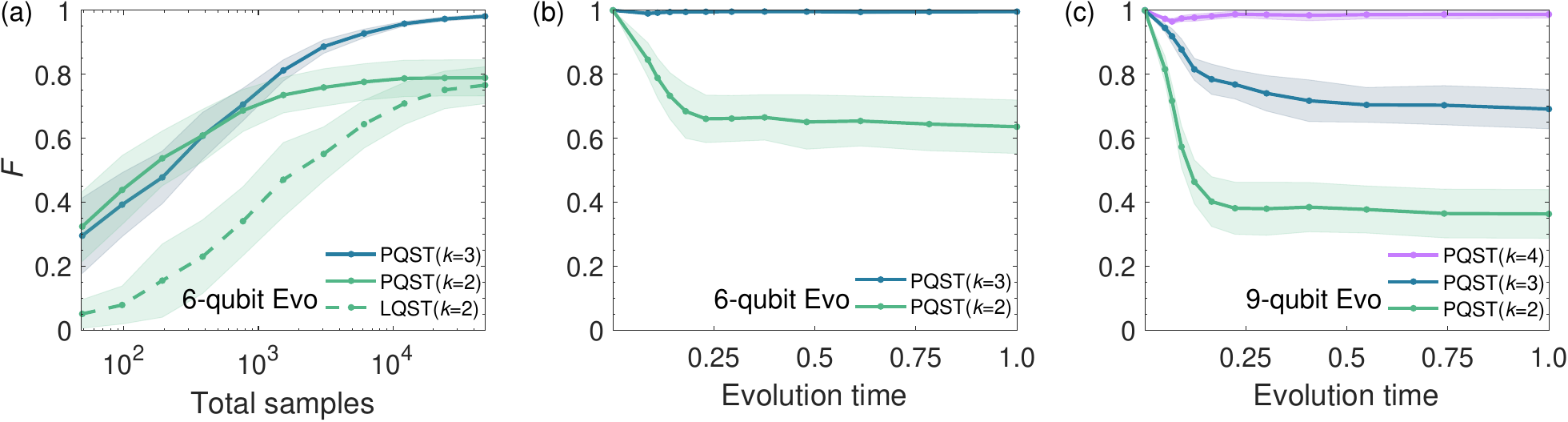}
	\caption{
		\textbf{%The numerical reconstruction results of PQST on dynamical states of FC Hamiltonian models
  The reconstruction results of the dynamical states}. (a) denotes the tomographic results of 6-qubit evolution states at evolution time $t=0.1$ s with different total measurement samples $M_{\text{tot}}$. (b) denotes the tomographic results of the 6-qubit evolution states in different evolution times and (c) denotes the results for 9-qubit evolution states.}
	\label{fig:sim_6q}
\end{figure}

To further validate the generality of our PQST method, we numerically reconstructed the dynamical states under the fully connected (FC) Hamiltonian  \(\mathcal{H}_{\mathrm{FC}}\) shown in Eq. (\ref{eq:HFC}).
 The FC Hamiltonian \(\mathcal{H}_{\mathrm{FC}}\) drives the quantum system from the initial state \(|\psi_0\rangle = |0\rangle^{\otimes N}\) to the final state \(|\psi_t\rangle = e^{-i\mathcal{H}_{\mathrm{FC}}t}|\psi_0\rangle\).
 We obtained measurement outcomes on final states via the numerical Monte Carlo simulations, considering various times and different $k$-qubit RDMs. Then we reconstruct these dynamical states using the loss function $f_{\text{MSE}}$. The fidelities of the reconstructed states are shown in Fig.~\ref{fig:sim_6q}.

 By applying PQST and LQST to reconstruct the dynamical states at a fixed evolution time \(t=0.1\) s, we evaluated these the efficiency of these methods in different measurement samples. As illustrated in Fig.~\ref{fig:sim_6q}(a), we plotted the fidelity achieved via each method as a function of the total samples $M_{\text{tot}}$.
This comparison demonstrates that PQST is more efficient than LQST in terms of measurement costs, as it achieves the same fidelity with fewer measurements.
However, despite increasing the total measurement samples, the fidelities of reconstructed density matrix for both PQST $(k=2)$ and LQST $(k=2)$ are below 0.9. This indicates that two-qubit RDMs are insufficient to fully characterize this quantum state. The near-perfect fidelity is achieved using three-qubit RDMs in PQST.

We also examined the fidelities of the reconstructed density matrices at various evolution time using PQST method. A large enough number of total samples were used in this discussion. As shown in Fig.~\ref{fig:sim_6q}(b), PQST with $k = 3$ can accurately reconstruct the 6-qubit evolution states under $\mathcal{H}_{\mathrm{FC}}$ for $t\in [0, 1]$. However, for the 9-qubit case in Fig.~\ref{fig:sim_6q}(c), the high-fidelity reconstruction requires $k = 4$. This is because, as the system evolves, the interactions encoded in the Hamiltonian spread across the entire system.

\newpage

\subsection{The further experimental results}
 
\begin{figure}[H]
	\centering
	\includegraphics[width=130mm]{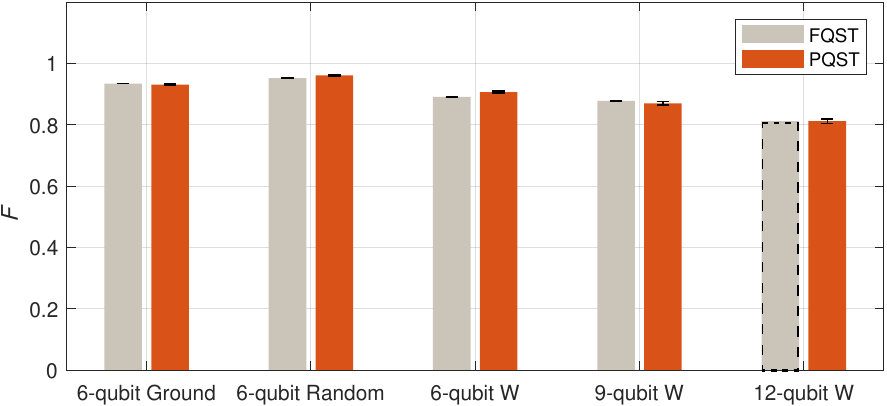}
    \caption{\textbf{The fidelities of the reconstructed states using both PQST and FQST methods with the theoretically prepared states are presented for different types of states. }The PQST results are obtained based on the last point of the PQST curves in Fig. 3(a-c) and Fig. 4(c) of the main text. The fidelity of the 12-qubit FQST with the theoretically prepared states is not shown (outlined with a dashed frame) due to its impractical time cost.} 
	\label{fig:fidelity_ideal}
\end{figure}

\begin{figure}[H]
	\centering
	\includegraphics[width=150mm]{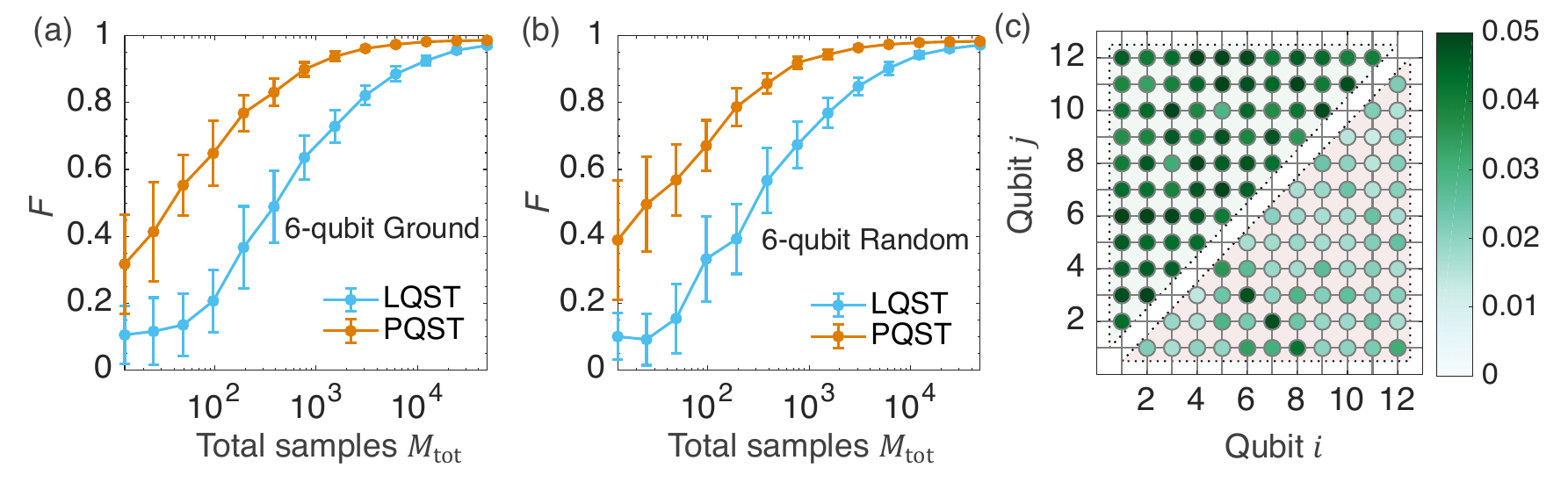}
	\caption{(a-b) The fidelities of the reconstructed density matrices with those obtained from FQST under different total samples $M_{\text{tot}}$ for 6-qubit ground states and random states, respectively. Here, we consider only two-qubit RDMs and use the loss function $f_{\text{MSE}}$ in LQST and PQST. (c) The error bars of the measured two-qubit correlation $\langle \sigma^{(i)}_y\sigma^{(j)}_y\rangle-\langle \sigma^{(i)}_y\rangle\langle\sigma^{(j)}_y\rangle$ using local (top left) and parallel (bottom right) measurements. The respective total samples $M_{\text{tot}}$ are $594M$ and $27M$ with $M=500$, respectively. The error bars are obtained by calculating the uncertainty when we repeat the sampling process for a given number of measurement samples. }
	\label{fig:ground}
\end{figure}

In the main text, we provide a detailed comparison of fidelities between the density matrices reconstructed using PQST and the ones obtained via FQST. Here, we further extend this analysis by presenting the fidelities between the reconstructed states obtained through FQST and PQST and their corresponding theoretically prepared states. Figure ~\ref{fig:fidelity_ideal} presents these fidelities, showing that the fidelity of PQST matches well with that of FQST and also exhibits small fidelity uncertainties, despite PQST requiring far fewer measurement samples than FQST.  In the main text, we also prepare and reconstruct ground states of 6-qubit FC Hamiltonians and random states, presenting the fidelities between the reconstructed states using PQST and FQST. The fidelities obtained through the LQST method are shown in Fig.~\ref{fig:ground}(a-b). The results also indicate that PQST is more sample-efficient than LQST, with smaller fidelity uncertainties under the same number of measurement samples. The uncertainty primarily arises from fluctuations in the projection statistics due to the finite sample size. Besides, we demonstrate the strong power of parallel measurements in measuring two-qubit correlators on the 12-qubit W state, as shown in Fig. 4(a) of the main text. Here, the corresponding error bars are provided in Fig.~\ref{fig:ground}(c). Clearly, the parallel measurement method not only reduces measurement costs but also results in smaller uncertainties.

% \begin{figure}[H]
% 	\centering
% 	\includegraphics[width=130mm]{Figs/fig_theo_f.pdf}
%     \caption{{\color{red}\textbf{The fidelities of the reconstructed states using both PQST and FQST methods with the theoretically prepared states are presented for different types of states. }The PQST method uses a total $M_{\text{tot}}=5\times 10^4$, $2\times 10^5$, and $5\times 10^5$ samples for the $6$-, $9$-, $12$-qubit cases, respectively. FQST employs  $M_{\text{tot}}=3^N\times 10^4$ samples for $N=6$ and $N=9$. The fidelity of the 12-qubit FQST with the theoretically prepared states is not shown (indicated by a dashed bar) because we did not experimentally perform FQST on a 12-qubit state.} }
% 	\label{fig:fidelity_ideal}
% \end{figure}

\begin{figure}[H]
	\centering
	\includegraphics[width=135mm]{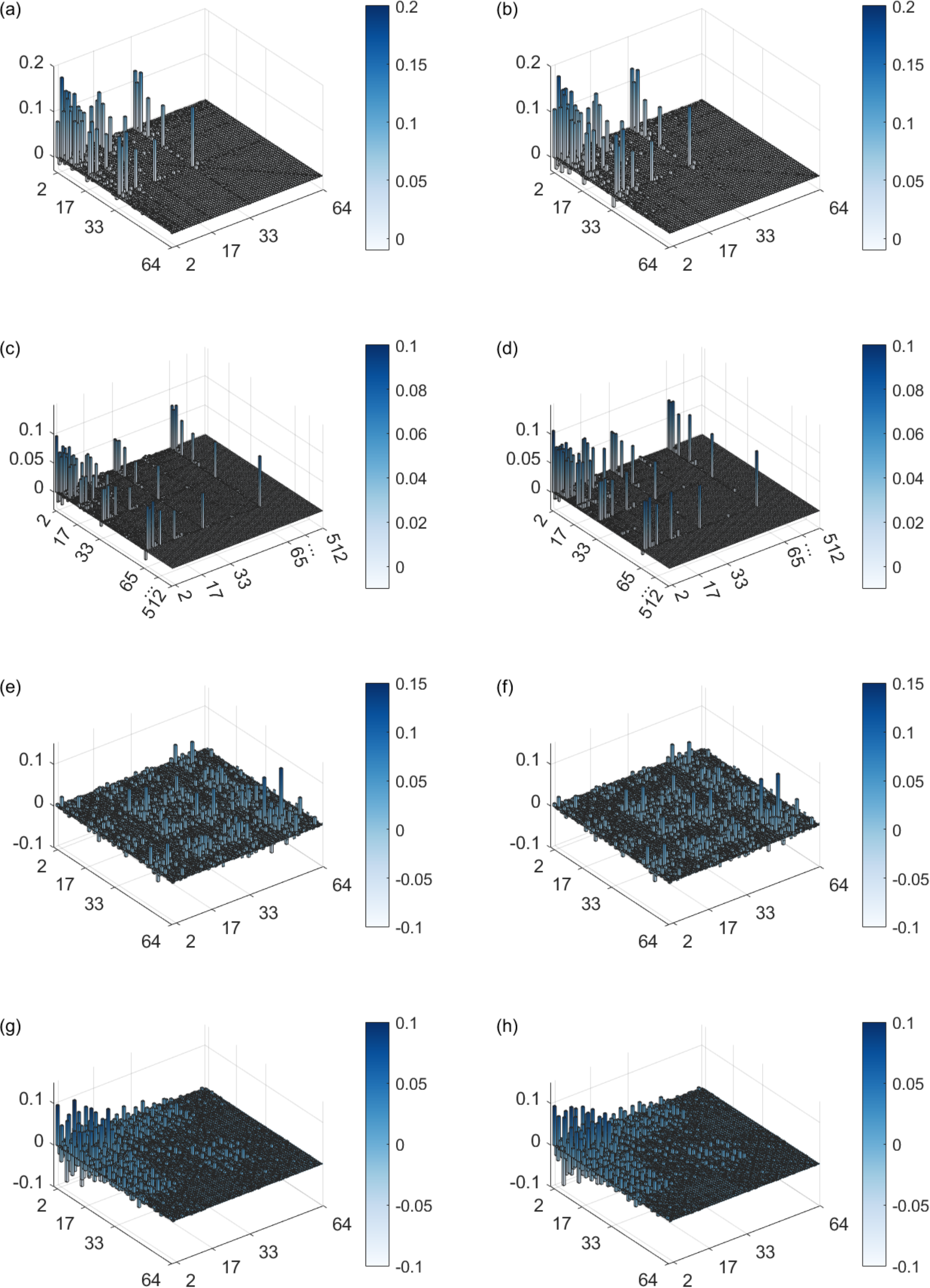}
	\caption{\textbf{The reconstructed density matrices using FQST and PQST.}  (a-b) display the reconstructed density matrices for the 6-qubit W state. (c-d) show the results for the 9-qubit W state. (e-f) depict the reconstructed density matrices for the 6-qubit ground state. (g-h) illustrate the results for the 6-qubit random state. The left column presents the results from PQST, while the right column displays those from FQST. The \(x\) and \(y\) axes represent the computational basis of the Hilbert state space. For the 9-qubit state, due to its size, we only display the left and right parts of the \(2^9 \times 2^9\) density matrix, as the entire matrix is not feasible.}
	\label{fig:exp_st_wst}
\end{figure}

Additionally, we provide a direct comparison of the reconstructed density matrices obtained through both FQST and PQST.
Figure ~\ref{fig:exp_st_wst}(a-b) show the mixed density matrices for the 6-qubit W state, reconstructed using PQST ($f^{k=3}_{\text{MLE}}$), which measures ${D^{(6,3)}_{\text{PQST}}=75}$ parallel observables, and FQST, which measures $3^6=729$ observables, respectively. 
Figure ~\ref{fig:exp_st_wst}(c-d) present similar comparisons for the 9-qubit W state, reconstructed with PQST measuring ${D^{(9,3)}_{\text{PQST}}=99}$ parallel observables and FQST measures $3^9=19683$ observables. Figure ~\ref{fig:exp_st_wst}(e-f) and (g-h) depict the reconstructed density matrices for the 6-qubit ground state and the random state, respectively. Here, PQST only measures ${D^{(6,2)}_{\text{PQST}}=21}$ parallel observables for the 6-qubit ground states and the random state in experiments. The close agreement between these reconstructions demonstrate the superiority for quantum state tomography.

\begin{figure}[H]
	\centering
	\includegraphics[width=180mm]{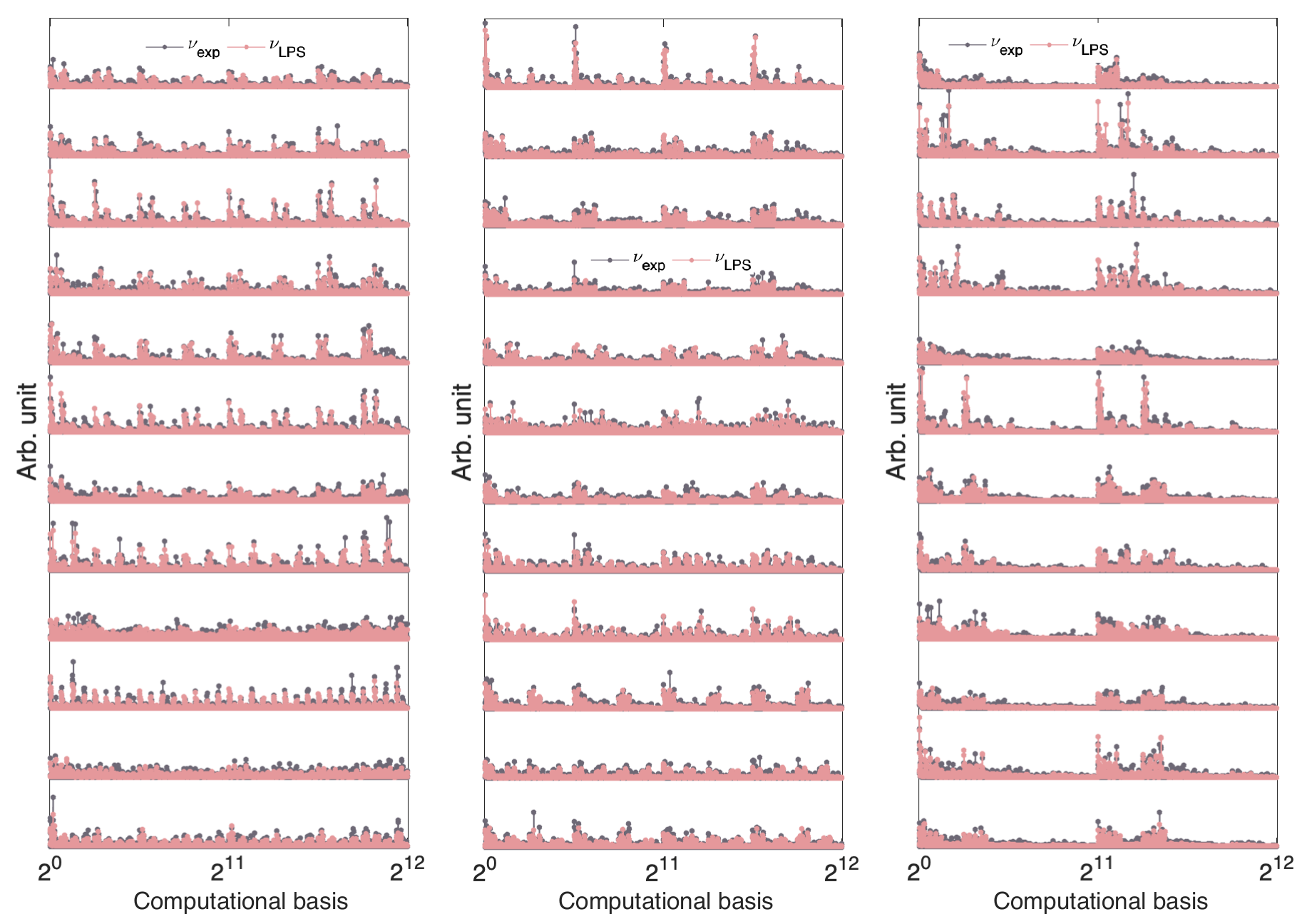}
	\caption{\textbf{Comparison of projection probability distributions from the reconstructed density matrix via PQST.} We randomly present comparison results for 36 out of 200 additional observables. The 12-qubit mixed density matrix, reconstructed using PQST, is based on measurements of just 243 observables, whereas FQST will require measurements of \(3^{12} = 531441\) observables. This highlights the efficiency of PQST in handling large-scale quantum state tomography with significantly fewer measurements.}
	\label{12q}
\end{figure}

Due to the prohibitively long experimental time (approximately 60 days) required for FQST, we choose an alternative validation of the PQST method for the 12-qubit case. We randomly selected 200 additional observables not measured in PQST and determined their corresponding measurement outcomes experimentally. For every observable, we reshape these measurement projection distributions into vectors $\bf{\nu}_{\text{exp}}$ and compare them with the vectors $\bf{\nu}_{\text{LPS}}$ obtained from the reconstructed density matrix using PQST.
In the main text, we presented the cosine similarity between $\bf{\nu}_{\text{exp}}$ and $\bf{\nu}_{\text{LPS}}$ for all 200 additional observables, and we visualized the projection distributions of $\bf{\nu}_{\text{exp}}$ and $\bf{\nu}_{\text{LPS}}$ for one of observables. As illustrated in Fig.~\ref{12q}, we provided $\bf{\nu}_{\text{exp}}$ and $\bf{\nu}_{\text{LPS}}$ for other 36 observables. The good agreement between $\bf{\nu}_{\text{exp}}$ and $\bf{\nu}_{\text{LPS}}$ across all observables indicates that the density matrix reconstructed using PQST accurately represents the experimentally prepared state.

\newpage

\section{LIST OF ABBREVIATIONS}
 \begin{tabular}{p{2cm}l}
QST & Quantum state tomography \\
LQST & Quantum state tomography via local measurements\\
PQST & Quantum state tomography via parallel measurements\\
FQST & Full quantum state tomography\\
QOT & Quantum overlapping tomography\\
MSE & Mean squared error\\
MLE & Maximum likelihood estimation\\
LPS & Locally purified state\\
RDM & Reduced density matrix\\
$\binom{N}{k}$ & The number of $k$-combinations of an $N$-element set, $\binom{N}{k} = \frac{N!}{k!(N-k)!}$\\
$D_{\text{LQST}}$ & The number of local observables within the LQST framework\\
$D_{\text{PQST}}$ & The number of parallel observables within the PQST framework\\
$M$ & The number of measurement samples on each measurement observable\\
$M_{\text{tot}}$ & The total number of measurement samples on whole measurement observables\\
$\mathcal{L}_i$ & The $i$-th  local observable\\
$\eta_i$ & The expectation values of the $i$-th local observables $\mathcal{L}_i$\\
$\ket{\gamma^j_s}$ &  The outcome state of the $s$-th shot on the $j$-th parallel observable\\ 
$\ket{W_{N}}$ & The $N$-qubit W state\\
$\rho_{\text{LPS}}$ & The reconstructed density matrix using the LPS method\\
$\rho_{\text{FQST}}$ & The reconstructed density matrix via FQST from experiments\\
$\bf{\nu}_{\text{exp}}$ & The projection distributions onto the eigenvectors of parallel observables from experiments\\
$\bf{\nu}_{\text{LPS}}$ & The projection distributions onto the eigenvectors of parallel observables from the LPS method\\
\end{tabular}

% \bibliography{SupInfo_refer.bib}

% \bibliographystyle{PRA-WithTitle.bst}
